\documentclass[aps,prb,superscriptaddress,reprint,twocolumn]{revtex4-2}

\usepackage{amsmath,amssymb,bm,bbm,braket,mathrsfs,mathtools}
\usepackage{multirow}
\usepackage{hyperref}
\usepackage[dvipdfmx]{graphicx}
\usepackage{color}
\usepackage[table]{xcolor}
\usepackage{array}

\def\sgn{\mathop{\textrm{sgn}}}
\def\Tr{\mathop{\textrm{Tr}}}

\newcommand{\hattimes}{\mathbin{\hat{\otimes}}}
\newcommand{\bartimes}{\mathbin{\bar{\otimes}}}
\newcommand{\eq}[1]{Eq.~(\ref{#1})} 

\begin{document}

\allowdisplaybreaks

\title{Bogoliubov Fermi surfaces from pairing of emergent $j=3/2$ fermions\\
on the pyrochlore lattice}
\author{Shingo Kobayashi}
\affiliation{RIKEN Center for Emergent Matter Science, Wako, Saitama 351-0198, Japan}
\author{Ankita Bhattacharya}
\affiliation{Institute of Theoretical Physics, Technische
  Universit\"at Dresden, 01062 Dresden, Germany}
\author{Carsten Timm}
\affiliation{Institute of Theoretical Physics, Technische
  Universit\"at Dresden, 01062 Dresden, Germany}
\affiliation{W\"urzburg-Dresden Cluster of Excellence ct.qmat,
  Technische Universit\"at Dresden, 01062 Dresden, Germany}
\author{P. M. R. Brydon}
\affiliation{Department of Physics and MacDiarmid Institute, University of Otago, PO Box 56, Dunedin 9054, New Zealand}

\date{\today}

\begin{abstract}
We examine the appearance of superconductivity in the strong-coupling
limit of the Hubbard model on the pyrochlore lattice. We focus upon
the limit of half filling, where the normal-state band structure realizes
a $j=3/2$ semimetal. Introducing doping, we show that the
pairing is favored in a $J=2$ quintet $E_g$ state. The attractive interaction in this channel relies on the fact that $E_g$ pairing on the pyrochlore lattice avoids the detrimental on-site repulsion.
Our calculations show that a time-reversal symmetry-breaking superconducting phase is
favored, which displays Bogoliubov Fermi surfaces.
\end{abstract}

\maketitle

\section{Introduction}

The physics of pyrochlore systems such as the iridate compounds
$R_2$Ir$_2$O$_7$ ($R$ is a rare-earth element) has attracted much
attention over the past
decade~\cite{Pesin2010,Wan2011,Witczak2012,WKGK2013,Lee2013,Savary2014,Bzdusek2015,Goswami2017,Laurell2017,BMT18}. These
materials are characterized by the interplay of strong electronic
correlations and strong spin-orbit coupling~\cite{WKCKB2014}, which is
predicted to yield a variety of exotic correlated states, such as spin
liquids~\cite{Pesin2010} and magnetically-ordered states with
nontrivial
topology~\cite{Wan2011,Witczak2012,WKGK2013,Lee2013,Savary2014,Bzdusek2015,Goswami2017,Laurell2017,BMT18}. 
The pyrochlore crystal structure of these materials is
characterized by a lattice of corner-sharing tetrahedra composed of
Ir$^{4+}$ ions, with the
low-energy electronic states deriving from the spin-orbit-split
$J_{\text{eff}}=1/2$ doublet of the $t_{2g}$ manifold of the Ir $5d$
orbitals~\cite{Pesin2010}. Due to the cubic structure of the pyrochlore lattice,
the low-energy Bloch states deriving from
the $J_{\text{eff}}=1/2$ doublets of the four Ir ions in each unit
cell can possess a nontrivial emergent $j=3/2$ angular
momentum. This emergent angular momentum describes states near
quadratic band-touchings at the Brillouin-zone center, which have been
observed in a number of pyrochlore iridates~\cite{Kondo2015,Nakayama2016}.

Fermionic systems with $j=3/2$ have been proposed to host a
number of exotic ordered phases and possibly non-Fermi-liquid
behavior~\cite{Savary2014,Boettcher2017,Goswami2017}. In particular,
the allowed superconducting states are much enriched: in addition to
pairing in a spin-singlet ($J=0$) or triplet ($J=1$) channel,
pairing in quintet ($J=2$) or septet ($J=3$) states is also
allowed~\cite{BWW16}. These higher spin states can display gap
functions with remarkable nodal structures, e.g., Bogoliubov Fermi
surfaces (BFSs)~\cite{ABT17,BAM18,TiB21,Kim2021,Dutta2021} or Dirac superconductors with
quadratic or cubic nodal dispersions~\cite{Venderbos2018}. So far,
however, these states have mostly been discussed in terms of the
effective Luttinger model valid near the quadratic band-touching
point~\cite{BWW16,RGF19,TGWK20,SMR21,BoH18}, whereas theories of
unconventional superconductors are more typically formulated in terms
of tight-binding models with local interactions. Using the
latter perspective,
Laurell and Fiete \cite{Laurell2017} have studied superconductivity
in a quasi-two-dimensional model of a pyrochlore lattice, but the
breaking of cubic symmetry implies that the
quasiparticles do not have $j=3/2$ character.
 
In this paper we motivate the pyrochlore lattice as a minimal
tight-binding model in which to study the superconductivity of
fermions with an emergent $j=3/2$ effective angular
momentum. Including an on-site Hubbard repulsion $U$, we derive the
pairing interaction in the strong-coupling limit. We find that the
dominant pairing instability will be the extended \textit{s}-wave
$E_g$ pairing channel corresponding to $J=2$ quintet pairing,
which is likely to realize a time-reversal-symmetry-breaking state
with BFSs. 

Our paper is organized as follows: In Sec.~\ref{sec:pyrochlore}, we introduce the tight-binding model of the pyrochlore lattice and determine the parameter regime where the $j=3/2$ fermionic quasiparticles are the low-energy excitations at half filling. The parameters of the effective Luttinger model are obtained in terms of the tight-binding parameters. In Sec.~\ref{sec:pairing}, we postulate a general interaction Hamiltonian for our tight-binding model, including both on-site and nearest-neighbor interactions. We project this interaction onto the low-energy states and decouple it in the Cooper channel, restricting our attention to the states with nonzero pairing amplitude at the Brillouin-zone center, namely the singlet $A_{1g}$ state and the quintet $E_g$ and $T_{2g}$ states. Specializing to the strong-coupling limit, where the nearest-neighbor interaction potentials perturbatively arise from virtual hopping events, we argue in Sec.~\ref{sec:A1gT2g} that the effective pairing interaction is repulsive in the $A_{1g}$ and $T_{2g}$ channels. In contrast, the pairing interaction is attractive in the $E_g$ channel, which we show in Sec.~\ref{sec:Eg} is generically realized in a time-reversal-symmetry-breaking state supporting BFSs.

\section{$j=3/2$ fermions on the pyrochlore lattice}
\label{sec:pyrochlore}

The fundamental structural feature of the pyrochlore lattice are
corner-sharing tetrahedra. The tetrahedra which do not directly touch
one another form an fcc lattice. Taking the centers of these tetrahedra
as the lattice points, the basis vectors for the four atoms are given
by 
\begin{align}
\mathbf{b}_1 &= \frac{a}{4} \left(-\frac12,-\frac12,-\frac12\right) , \\
\mathbf{b}_2 &= \frac{a}{4} \left(\frac12,\frac12,-\frac12\right) , \\
\mathbf{b}_3 &= \frac{a}{4} \left(-\frac12,\frac12,\frac12\right) , \\
\mathbf{b}_4 &= \frac{a}{4} \left(\frac12,-\frac12,\frac12\right) ,
\end{align}
where $a$ is the lattice constant of the conventional fcc unit cell.

The standard electronic model for the pyrochlore iridates is a
tight-binding model extending up to next-nea\-rest neighbors for
Ir $J_{\text{eff}}=1/2$ doublets at each pyrochlore
site~\cite{WKGK2013}. For simplicity, henceforth we label these
doublets by a spin degree of freedom $\{\uparrow,\downarrow\}$. The noninteracting model is described by the Hamiltonian 
\begin{align}
  \label{eq:H0}
  H =&  \sum_{\langle ij\rangle}c^{\dagger}_{i}\left (t_1 + it_2{\bf d}_{ij}\cdot\bm{\sigma}\right)c^{}_{j} \notag \\
& + \sum_{\langle\langle ij\rangle\rangle}c^{\dagger}_{i}\left (t^\prime_1 + i\left[t_2^\prime{\bf R}_{ij} + t_3^\prime{\bf D}_{ij}\right]\cdot\bm{\sigma}\right)c^{}_{j} ,
\end{align}
where $c^{}_i = (c_{i,\uparrow},c_{i,\downarrow})^T$ is the spinor of
creation and annihilation operators for the doublet states at site $i$, $\bm{\sigma}$ is the vector of Pauli matrices, and the vectors appearing in~\eq{eq:H0} are defined as~\cite{BMT18}
\begin{align}
{\bf d}_{ij} &= 2{\bf b}_{i}\times{\bf b}_{j} , \\
{\bf R}_{ij} &= ({\bf b}_i - {\bf b}_k)\times ({\bf b}_k - {\bf b}_j) , \\
{\bf D}_{ij} &= {\bf d}_{ik}\times{\bf d}_{kj} ,
\end{align}
where $k$ is a common nearest neighbor of sites $i$ and $j$.
In the context of the iridates, the hopping integrals appearing
in Eq.~(\ref{eq:H0}) can be expressed
in terms of direct iridium-iridium hopping via $\sigma$ and $\pi$
bonds ($t_\sigma$, $t_\pi$, $t_\sigma^\prime$, $t_\pi^\prime$) and
also indirect hopping via oxygen ions ($t_O$). The Slater-Koster
method then predicts~\cite{WKGK2013}
\begin{align}
t_1 &= \frac{130}{243}\, t_O + \frac{17}{324}\, t_\sigma - \frac{79}{243}t_\pi , \\
t_2 &= \frac{28}{243}\, t_O + \frac{15}{243}\, t_\sigma - \frac{40}{243}\, t_\pi , \\
t_1^\prime &= \frac{233}{2916}\, t_\sigma^\prime - \frac{407}{2187}\, t_\pi^\prime , \\
t_2^\prime &= \frac{1}{1458}\, t_\sigma^\prime + \frac{220}{2187}\, t_\pi^\prime , \\
t_3^\prime &= \frac{25}{1458}\, t_\sigma^\prime + \frac{460}{2187}\, t_\pi^\prime .
\end{align}

As mentioned in the introduction,
the pyrochlore structure naturally gives rise to $j=3/2$ fermionic
excitations. This is most easily understood by considering the
electronic structure of an isolated tetrahedron with an Ir ion with a $J_{\text{eff}}=1/2$ doublet at each vertex. The
orbital component of the electron wavefunctions for this
four-site cluster can be decomposed into an \textit{s}-wave-like
($A_1$ irrep of the point group $T_d$ of a tetrahedron) and three
\textit{p}-wave-like ($T_2$ irrep) wavefunctions. Spin-orbit coupling
splits the electronic states of the isolated tetrahedron into two
$j=1/2$ doublets and a $j=3/2$ quartet~\cite{Park2020}. For the full
pyrochlore lattice, this emergent electronic structure persists close to the $\Gamma$ point. 
In the following, we will focus on the case where the low-energy
excitations are due only to the $j=3/2$ fermions. 
In particular, this is possible if the $j=3/2$ bands are half filled, in which case
a semimetallic state with a quadratic band-touching
point may be realized. This is of 
special interest as combining the half-filling condition with
interactions raises the possibility of strongly-correlated $j=3/2$
fermions~\cite{BoH16}.

\begin{figure}
(a)\hspace*{-1ex}\includegraphics[width=0.47\columnwidth]{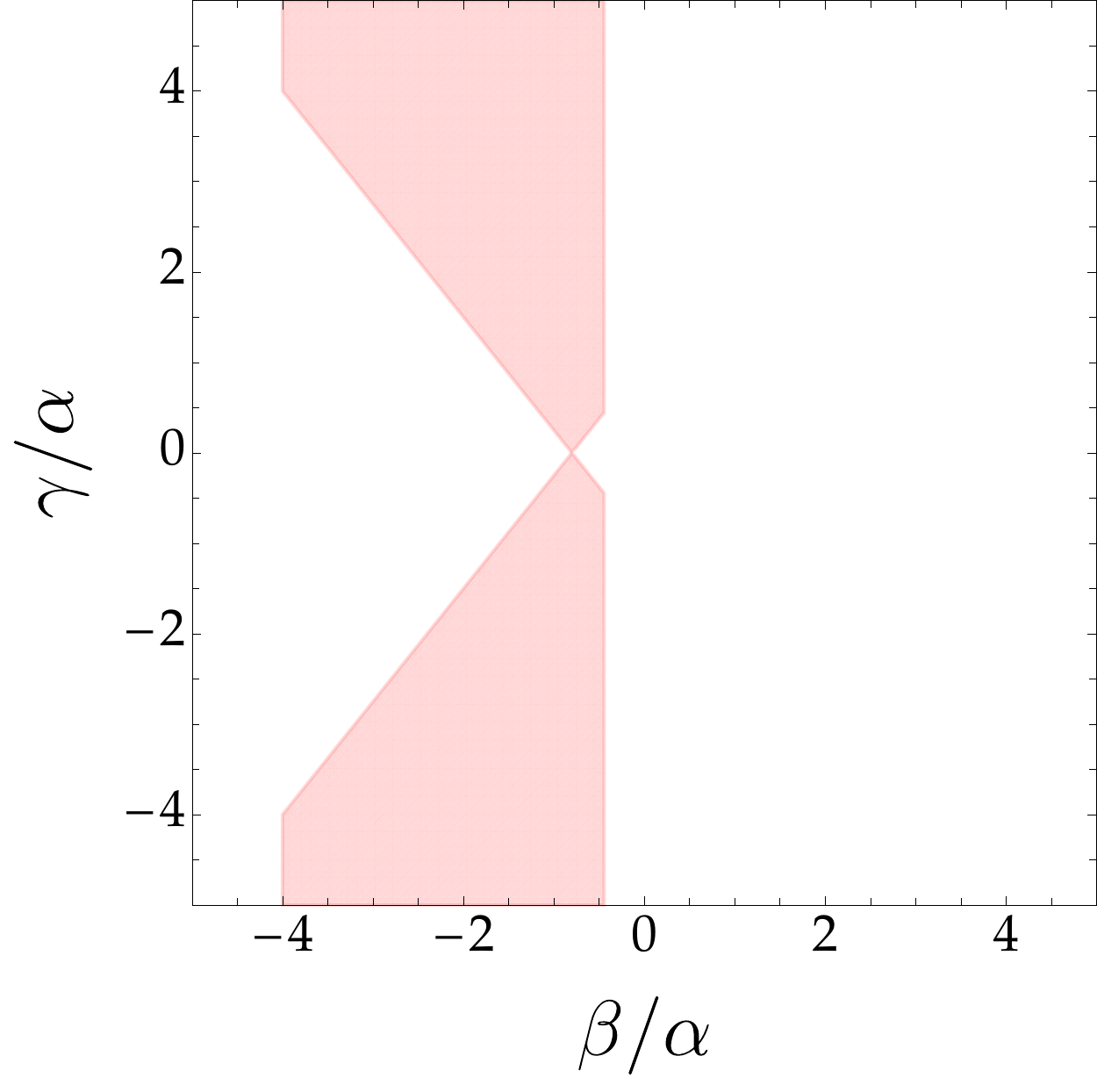}~%
(b)\hspace*{-1ex}\includegraphics[width=0.47\columnwidth]{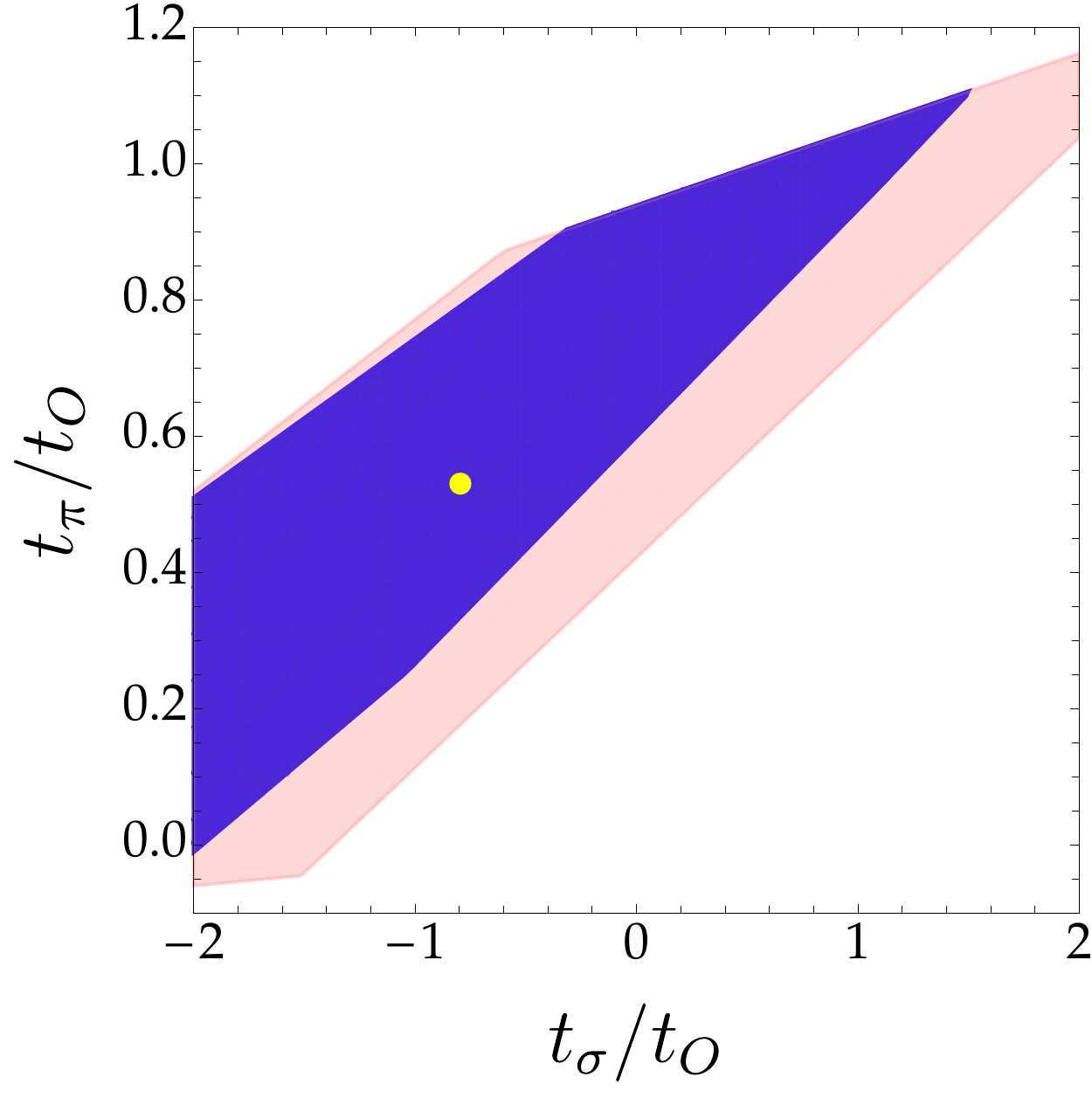}\\
(c)\hspace*{-1ex}\includegraphics[width=0.97\columnwidth]{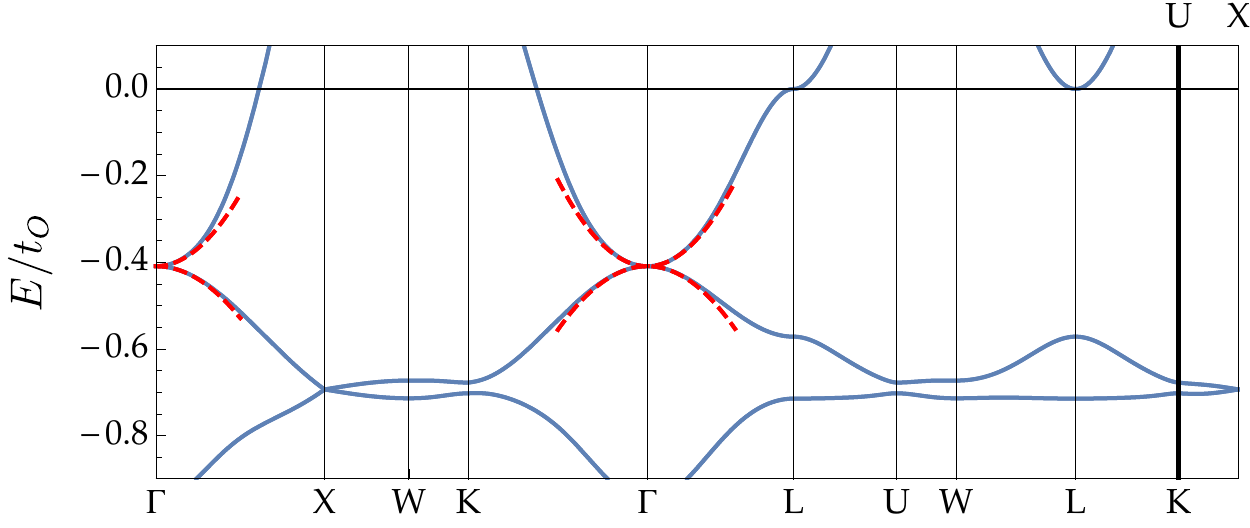}

\caption{(Color online) (a) Range of parameters $\beta$ and $\gamma$ in the
  Luttinger-Kohn Hamiltonian where a $j=3/2$
  semimetal is realized (shaded). (b) Range of parameters $t_\sigma$ and $t_\pi$
  where a $j=3/2$ semimetal state
  is realized for the pyrochlore lattice:
  The pink (light gray) region indicates
  where a semimetallic quadratic band touching is found at the
  $\Gamma$ point and is obtained by mapping the region in panel (a) to
  the pyrochlore lattice using
   Eqs.\ (\ref{eq:LKmapping1})--(\ref{eq:LKmapping3}). Within the enclosed
  dark blue (dark gray) region, there are no other states elsewhere in the
  Brillouin zone at the same energy as the quadratic band-touching
  point. (c) Typical band structure along high-symmetry directions
  showing the presence of a $j=3/2$ semimetal state. The parameter
  choice ($t_\sigma=-0.795t_O$, $t_\pi = 0.53t_O$) is indicated by the yellow dot in panel (b).
  The dashed red lines show the low-energy bands in the equivalent Luttinger-Kohn model.}
\label{fig:SM_LK}
\end{figure}

To quadratic order in momentum, the $j=3/2$ excitations close to the $\Gamma$ point are described by an effective Luttinger-Kohn model with Hamiltonian matrix
\begin{equation}
H_{\text{LK}}(\mathbf{k}) = \alpha |\mathbf{k}|^2 \hat{\mathbbm{1}} +
\beta \sum_{\mu}k_\mu^2\hat{J}_\mu^2 + \gamma \sum_{\mu \neq \nu}k_\mu k_\nu \hat{J}_\mu \hat{J}_\nu ,
\label{eq:HLK}
\end{equation}
where $\alpha$, $\beta$, and $\gamma$ are constants, $\hat{\mathbbm{1}}$ is the $4\times 4$ identity matrix,
and $\hat{J}_\mu$, $\mu=x,y,z$, are the $j=3/2$ angular-momentum matrices; see Appendix~\ref{sec:derivation_hlk} for a detailed derivation. 
The two distinct eigenvalues of the Luttinger-Kohn model are
\begin{align}\label{eq:ELK}
E_{\pm,\mathbf{k}} &= \left(\alpha + \frac{5}{4}\beta\right)|\mathbf{k}|^2  \notag \\
& \quad{}\pm \sqrt{\beta^2\sum_{\mu}k_\mu^4 + (3\gamma^2-\beta^2)\sum_{\mu<\nu}k_\mu^2k_\nu^2} .
\end{align}
 Both are twofold degenerate.
By considering the dispersion along the $[100]$ and
$[111]$ directions, the conditions for the bands
  to have opposite curvature, and thus for
a semimetal, are
\begin{align}
\sgn\left[\left(\alpha + \frac{1}{4}\,\beta\right)\left(\alpha +
    \frac{9}{4}\,\beta\right)\right] &= -1 , \label{eq:conds1} \\
\sgn\left[\left(\alpha + \frac{5}{4}\,\beta + |\gamma|\right)\left(\alpha +
    \frac{5}{4}\,\beta - |\gamma|\right)\right] &= -1 . \label{eq:conds2}
\end{align}
In Fig.~\ref{fig:SM_LK}(a), we plot the region in parameter space where
these conditions are satisfied. 

Projecting the pyrochlore Hamiltonian onto the $j=3/2$
subspace, we recover the Luttinger-Kohn model with the coefficients~\cite{Goswami2017}
\begin{align} 
\alpha + \frac{5}{4}\,\beta &= \frac{2}{3}\left(t_1 + 2t_2
  + 6t_{1}^\prime - 12t_2^\prime - 12t_3^\prime\right) , 
\label{eq:LKmapping1} \\
\sqrt{3}\,\gamma &= -\frac{2}{\sqrt{3}}\left(t_1 + 2t_2 - 2t_1^\prime
  - 4t_2^\prime + 12t_3^\prime\right) , \\
\frac{\sqrt{3}}{2}\,\beta &= -\frac{1}{\sqrt{3}}\left(t_1 -t_2
  - 6t_1^\prime - 6t_2^\prime - 6t_3^\prime \right) .
\label{eq:LKmapping3}
\end{align}
The five distinct Slater-Koster hopping integrals would give a large
parameter space to explore. However, we shall 
follow convention and set $t_O=1$ as our reference and then further
impose that
\begin{equation}
t_\sigma^\prime = 0.08\, t_\sigma, \qquad t_\pi^\prime = 0.08\, t_\pi .
\end{equation}
Thus, we shall regard $t_\sigma$ and $t_\pi$ as free parameters. The
parameter range in which the conditions Eqs.\ (\ref{eq:conds1})
and (\ref{eq:conds2}) for a semimetallic quadratic band touching are
satisfied is shown by the pink region in
Fig.~\ref{fig:SM_LK}(b). In this region, the $j=3/2$ states also
lie between the two $j=1/2$ bands at the $\Gamma$ point, which is a
necessary condition for such a semimetallic state at half filling. 
Since the Luttinger-Kohn Hamiltonian is only valid close to the
$\Gamma$ point, however, it is possible that other states are present
elsewhere in the Brillouin zone at the same energy as the quadratic
band touching. Accounting for this shaves off some of the edges of the
region identified by the conditions Eqs.\ (\ref{eq:conds1}) and (\ref{eq:conds2}),
leaving the dark blue region in Fig.~\ref{fig:SM_LK}(b) as the
parameter range where the low-energy excitations result solely from the
quadratic band-touching point. Figure~\ref{fig:SM_LK}(c)
shows a comparison of the
tight-binding and Luttinger-Kohn model dispersions for the parameter
values corresponding to the yellow dot in
Fig.~\ref{fig:SM_LK}(b).

\section{Superconducting states}

\subsection{Pairing interactions}
\label{sec:pairing}

The most general interactions for spin-$1/2$ electrons consistent with the symmetry of the
pyrochlore lattice up to nearest neighbors have the form~\cite{Witczak2012,Laurell2017}
\begin{align}
H_{\text{int}} &= U_0\sum_{i}n_{i,\uparrow}n_{i,\downarrow} + U_1 \sum_{\langle i, j \rangle }n_{i}n_{j}
  + J \sum_{\langle i, j \rangle }\mathbf{S}_{i}\cdot\mathbf{S}_{j}\notag\\ 
&\quad{}+ D \sum_{\langle i, j \rangle }{\bf d}_{ij}\cdot(\mathbf{S}_{i}\times\mathbf{S}_{j})
  + \sum_{\langle i, j \rangle }\sum_{\mu,\nu}S^\mu_{i}\Gamma^{\mu \nu}_{ij}S^\nu_{j} , \label{eq:Hint}
\end{align}
where $n_{i,\sigma} \equiv c_{i,\sigma}^\dagger c_{i,\sigma}$ is a number operator and $\mathbf{S}_i$ is a spin operator with components $S_i^\mu \equiv \sum_{\sigma,\sigma'} c_{i,\sigma}^\dagger s^\mu_{\sigma\sigma'} c_{i,\sigma'}$, where $s^\mu = \sigma^\mu/2$ are the spin-$1/2$ matrices.
The first line of Eq.\ (\ref{eq:Hint}) contains an on-site Hubbard repulsion as well as nearest-neighbor charge-charge and Heisenberg interactions. The second line contains the Dzya\-lo\-shins\-ki-Mo\-ri\-ya interaction and the traceless symmetric interaction
\begin{equation}
\Gamma^{\mu\nu}_{ij}
  = d^\mu_{ij} d^\nu_{ij} (\Gamma_0\delta_{\mu\nu}+\Gamma_1[1-\delta_{\mu\nu}]) .
\end{equation}  
In performing the sum over nearest neighbors $\langle i, j \rangle$, we
count each bond once.

The nearest-neighbor interactions in $H_{\text{int}}$ naturally arise in the strong-coupling limit of the Hubbard model \cite{Lee2013,Laurell2017}. Ignoring next-nearest-neighbor hopping and assuming half filling and that the Hubbard energy $U_0$ greatly exceeds $t_1$ and $t_2$, we integrate out doubly occupied sites to obtain the effective interaction strengths
\begin{align}
 U_1 &= - \frac{2}{U_0}\, (t_1^2 + 2 t_2^2), \\
 J &= \frac{4}{U_0} \left( t_1^2 - \frac{2}{3} t_2^2\right), \\
 D &= \frac{8}{U_0}\, t_1 t_2, \\
 \Gamma_0 &= \frac{8 t_2^2}{3U_0}, \\
 \Gamma_1 &= - \frac{8 t_2^2}{U_0}.
\end{align}

To work in the more convenient $j=3/2$ subspace,
we project the interactions onto the low-energy states.
We express
the annihilation operator at site $a$ of tetrahedron $i$ in
terms of the local operators in the $j=3/2$ subspace,
\begin{equation}
c_{i,a,\sigma} \approx \sum_{\alpha=-3/2}^{3/2} u_{a,\sigma;\alpha}\, c_{i,\alpha} , \label{eq:transform}
\end{equation}
where the coefficients $u_{a,\sigma;\alpha}$ are obtained in
Appendix~\ref{sec:derivation_hlk}. 
Substituting Eq.\ (\ref{eq:transform}) into Eq.\ (\ref{eq:Hint}),
we obtain the effective interaction in the low-energy subspace,
\begin{align}
H_{\text{int}} &= \sum_{i}\sum_{\alpha,\alpha'}\sum_{\beta,\beta'}V_{\alpha \alpha';\beta \beta'}\,
  c^\dagger_{i,\alpha}c^{}_{i,\alpha'}c^\dagger_{i,\beta}c^{}_{i,\beta'} \notag \\
&\quad{}+\sum_{\langle i, j \rangle_{a a'}}\sum_{\alpha,\alpha'}\sum_{\beta,\beta'}
  [V_{a,a'}]_{\alpha \alpha';\beta \beta'}\,
  c^\dagger_{i,\alpha}c^{}_{i,\alpha'}c^\dagger_{j,\beta}c^{}_{j,\beta'},  \label{eq:j32int}
\end{align}
where the sum over $\langle i, j \rangle_{a a'}$ contains all
nearest-neighbor pairs of sites $a$, $a'$ on tetrahedra $i$,
$j$. The interaction potentials are given by
\begin{equation}
V_{\alpha \alpha';\beta \beta'} = U_0 \sum_{a} u_{a,\uparrow;\alpha}^{\ast} u_{a,\uparrow;\alpha'} u_{a,\downarrow;\beta}^{\ast} u_{a,\downarrow;\beta'}  \label{eq:Vonsite}
\end{equation}
for the on-site interaction and 
\begin{align}
[V_{a,a'}]_{\alpha \alpha';\beta \beta'} &= U_1 \sum_{\sigma_1,\sigma_2} u_{a,\sigma_1;\alpha}^{\ast} u_{a,\sigma_1;\alpha'} u_{a',\sigma_2;\beta}^{\ast} u_{a',\sigma_2;\beta'} \notag \\
&\quad {}+ \sum_{\sigma_1,\sigma_2, \sigma_3,\sigma_4}  \sum_{\mu, \nu} s_{\sigma_1 \sigma_2}^{\mu} s_{\sigma_3 \sigma_4}^{\nu}  \notag \\
&\qquad {}\times \Big( J \delta_{\mu\nu}+D \sum_{\rho} \epsilon^{\mu \nu \rho}  d_{aa'}^{\rho} +  \Gamma^{\mu \nu}_{aa'}\Big) \notag \\
&\qquad {}\times u_{a,\sigma_1;\alpha}^{\ast} u_{a,\sigma_2;\alpha'} u_{a',\sigma_3;\beta}^{\ast} u_{a',\sigma_4;\beta'} \label{eq:Vnn}
\end{align}
for the nearest-neighbor interactions. Here, $\epsilon^{\mu \nu \rho}$
is the Levi-Civita symbol. The lengthy explicit expressions for the
coefficients $u_{a,\sigma;\alpha}$ and the potentials $V$ and
$V_{a,a'}$ are relegated to
Appendix~\ref{sec:interaction_hlk}. 

We treat $H_{\text{int}}$ in \eq{eq:j32int} as an effective pairing interaction, which we eventually want to decouple in the Cooper channel. To that end, we decompose the interaction into the even-parity Cooper channels, using the generalized Fierz identity~\cite{Boettcher2017}
\begin{equation}
(\psi^{\dagger} N \psi)(\phi^{\dagger} M \phi) =\sum_{\hat{A},\hat{B}}
   f_{NM}(\hat{A},\hat{B})\, (\psi^{\dagger} \bar{A} \phi^{\dagger T})(\phi^{T} \bar{B}^{\dagger} \psi) , \label{eq:Fierz_identity_main}
\end{equation}
where
\begin{equation}
f_{NM}(\hat{A},\hat{B}) = \frac{1}{16}\, \Tr ( U_T^{\dagger} \hat{A}N\hat{B}U_T M^T ).
\end{equation}
The right-hand side of Eq.~(\ref{eq:Fierz_identity_main}) represents a pairing interaction, where $\bar{A} \equiv \hat{A} U_T $ and $U_T = \exp( i \pi \hat{J}_y)$ is the unitary part of the time-reversal operator. The matrices $\bar{A}$ describe the internal symmetry of the Cooper pairs in the $j=3/2$ space \cite{BAM18}. The six matrices compatible with even parity are listed in Table \ref{tab:os}, together with the corresponding irreps. $\psi$ and $\phi$ are field operators on a basis of $j=3/2$ fermions.

\begin{table}
\caption{ Internal symmetries of Cooper pairs allowed for even-parity pairing. The irreps of the point group $O_h$ and the pairing matrices are given.
The matrix $U_T = \exp( i \pi \hat{J}_y)$ is the unitary part of the time-reversal operator.
\label{tab:os}}
\begin{ruledtabular}
\begin{tabular}{cl}
irrep & pairing state\\\hline
$A_{1g}$ & $\bar{\mathbbm{1}}=U_T$\\
$E_{g}$ & $( \bar{E}_1, \bar{E}_2 ) = \frac{1}{\sqrt{3}}
  ( \hat{J}_x^2-\hat{J}_y^2,(2\hat{J}_z^2-\hat{J}_x^2-\hat{J}_y^2)/\sqrt{3} )\, U_T$\\
$T_{2g}$ & $( \bar{T}_1,\bar{T}_2,\bar{T}_3 ) = \frac{1}{\sqrt{3}}
  ( \{\hat{J}_y,\hat{J}_z\}, \{\hat{J}_z,\hat{J}_x\}, \{\hat{J}_x,\hat{J}_y\} )\, U_T$\\
\end{tabular}
\end{ruledtabular}
\end{table}

Transforming the interaction to momentum space and restricting ourselves to the pairing of electrons with opposite momenta, we write the pairing Hamiltonian as
\begin{align}
H_{\text{pair}} &= \frac{1}{2N}\sum_{\mathbf{k},\mathbf{k}'}\sum_{\alpha,\beta,\alpha',\beta'}[V_{\mathbf{k},\mathbf{k}'}]_{\alpha \beta;\alpha' \beta'} \nonumber \\
&\quad{}\times c^\dagger_{\mathbf{k},\alpha}c^\dagger_{-\mathbf{k},\beta}c^{}_{{\bf -k'},\alpha'}c^{}_{{\bf k'},\beta'}, \label{eq:h_pair}
\end{align}
where $N$ is the number of unit cells. The coupling strength contains contributions from the on-site interaction, Eq.\ (\ref{eq:Vonsite}), and from the nearest-neighbor interaction, Eq.\ (\ref{eq:Vnn}).
The undoped semimetal has the Fermi energy at the quadratic band-touching point and vanishing electronic density of states, and thus does not show superconductivity at weak coupling.
Upon doping the semimetal, a small Fermi surface will appear at the zone center. 
We hence restrict our study to the even-parity channel since odd-parity superconductivity has a vanishing pairing amplitude at the $\Gamma$ point and is thus typically weak at this small Fermi surface.

The decomposition into the even-parity Cooper channels is
  performed in Appendix \ref{sec:Cooperchannel_hlk}. Here, we focus on
  the limit of weak doping, i.e., $k_F \ll \pi/a$. In this limit, only
  those terms that remain nonzero for $\mathbf{k}\to 0$ are
  important. The resulting pairing interaction then reads as
\begin{align}
V_{\mathbf{k},\mathbf{k}'} &\approx \frac{U_0}{8}\, \bar{\mathbbm{1}} \bartimes \bar{\mathbbm{1}}'
  + \frac{U_0}{24}\, \vec{\bar{T}} \bartimes \vec{\bar{T}}' \notag \\
&\quad{} + \left(\frac{U_1}{18} - \frac{J}{216} - \frac{D}{27} - \frac{\Gamma_0}{108}+\frac{\Gamma_1}{54}\right)c_{A_{1g}} \bartimes c_{A_{1g}}' \notag \\
&\quad{} + \left(\frac{U_1}{9} - \frac{J}{108} + \frac{D}{27} - \frac{\Gamma_0}{54} - \frac{\Gamma_1}{54}\right)\vec{c}^{\, (E)}_{E_{g}} \bartimes \vec{c}^{\, (E)\,\prime}_{E_{g}} \notag\\
&\quad{} + \left(\frac{5U_1}{54} + \frac{J}{216} - \frac{D}{27} + \frac{\Gamma_0}{108} + \frac{\Gamma_1}{54}\right) \vec{c}^{\, (T)}_{T_{2g}} \bartimes \vec{c}^{\, (T)\,\prime}_{T_{2g}} , \label{eq:pair_int}
\end{align}
where we define the product $\bartimes$ to simplify the notation
such that for a given field operator $c_{\mathbf{k}}^T \equiv (c_{\mathbf{k},\frac{3}{2}},c_{\mathbf{k},\frac{1}{2}},c_{\mathbf{k},-\frac{1}{2}},c_{\mathbf{k},-\frac{3}{2}})$,
\begin{align}
&\sum_{\alpha ,\beta, \alpha', \beta'}  (\bar{A} \bartimes \bar{B})_{\alpha \beta;\alpha' \beta' } \, c_{\mathbf{k},\alpha}^{\dagger} c_{-\mathbf{k}, \beta}^{\dagger} c_{-\mathbf{k}',\alpha'} c_{\mathbf{k}', \beta'} \nonumber \\
&\qquad \equiv \sum_{\alpha ,\beta, \alpha', \beta'} \bar{A}_{\alpha \beta} \bar{B}_{\beta'  \alpha' }^{\ast} \, c_{\mathbf{k},\alpha}^{\dagger} c_{-\mathbf{k}, \beta}^{\dagger} c_{-\mathbf{k}',\alpha'} c_{\mathbf{k}', \beta'} \nonumber \\
&\qquad = \left(\sum_{\alpha ,\beta}c_{\mathbf{k},\alpha}^{\dagger} \bar{A}_{\alpha \beta} c_{-\mathbf{k}, \beta}^{\dagger} \right) \left( \sum_{\alpha' ,\beta'} c_{-\mathbf{k}',\alpha'} \bar{B}_{\beta'  \alpha' }^{\ast} c_{\mathbf{k}', \beta'} \right) \nonumber \\
&\qquad =(c_{\mathbf{k}}^{\dagger} \bar{A} c_{-\mathbf{k}}^{\dagger T})(c_{-\mathbf{k}'}^T \bar{B}^{\dagger} c_{\mathbf{k}'}). \label{eq:bartimes}
\end{align} 
 The first line of Eq.\ (\ref{eq:pair_int}) refers to on-site
  pairing, whereas the remaining terms result from nearest-neighbor
  interactions. The latter terms can be understood as extended
  \textit{s}-wave pairing, and contain the matrix-valued functions
\begin{align}
c_{A_{1g}} &= (c_x c_y + c_y c_z + c_z c_x)\, \bar{\mathbbm{1}} , \\
\vec{c}_{E_g}^{\,(E)} &= (c_x c_y + c_y c_z + c_z c_x)\,
  \big( \bar{E}_1, \bar{E}_2 \big) , \\
\vec{c}_{T_{2g}}^{\,(T)} &= (c_x c_y + c_y c_z + c_z c_x)\,
  \big( \bar{T}_1, \bar{T}_2, \bar{T}_3 \big) ,
\end{align}
with $c_\mu = \cos k_\mu a$. The prime signifies dependence on $\mathbf{k}'$. Full results are presented in Appendix~\ref{sec:Cooperchannel_hlk}.

Equation (\ref{eq:pair_int}) shows that on-site pairing in the $A_{1g}$ and $T_{2g}$
channels is penalized by the Hubbard interaction; in contrast, the
on-site $E_g$ pairing is immune to the Hubbard repulsion $U_0$, and 
there is no on-site interaction in this channel. This is
  a key result of our work. The nearest-neighbor interactions in Eq.\
  (\ref{eq:Hint}) lead to the momentum-dependent, extended
  \textit{s}-wave pairing terms in Eq.\ (\ref{eq:pair_int}). In the
  strong-coupling limit, the pairing potentials for extended
  \textit{s}-wave pairing are
\begin{align}
A_{1g}: &\quad -\frac{1}{378}\, \frac{(7t_1+8t_2)^2}{U_0} -
         \frac{121}{567}\, \frac{t_2^2}{U_0},
\label{eq:exts.A1g} \\
E_{g}: &\quad -\frac{1}{189}\, \frac{(7t_1+4t_2)^2}{U_0} -
         \frac{134}{567}\, \frac{t_2^2}{U_0}, \label{eq:exts.Eg}\\
T_{2g}: &\quad - \frac{1}{486}\, \frac{(9t_1+8t_2)^2}{U_0} -
         \frac{31}{243}\, \frac{t_2^2}{U_0}.
\label{eq:exts.T2g}
\end{align}
It is clear from this formulation that the interaction in the extended
\textit{s}-wave channels is always attractive.

\subsection{$A_{1g}$ and $T_{2g}$ channels} \label{sec:A1gT2g}

Both the on-site and extended \textit{s}-wave pairing potentials in the $A_{1g}$ and
$T_{2g}$ channels are nonzero. The states
will in general involve both components, e.g., in the case of the
$A_{1g}$ irrep we have $\Delta_{A_{1g}} = \Delta_o\, \bar{\mathbbm{1}} +
\Delta_{e}\, c_{A_{1g}}$. Following Ref.~\cite{OCS16}, the critical temperature of this mixed state is
obtained from the solution of the determinantal equation
\begin{equation}
\det \begin{pmatrix}
    \chi_{oo}-\frac{1}{g_o} & \chi_{oe}\\
    \chi_{oe} & \chi_{ee}-\frac{1}{g_e}
  \end{pmatrix} = 0 ,
\label{eq:det}
\end{equation}
where $g_o$ and $g_e$ are the interactions for the on-site and
extended \textit{s}-wave channels, respectively, and the generalized
superconducting susceptibilities are defined by
\begin{equation}
\chi_{ab} = {\cal N}_0 \int d\epsilon\,
  \frac{\tanh(\epsilon/k_BT)}{4\epsilon}\, \langle
  \Tr (\hat{\Delta}_{a}{\cal P}\hat{\Delta}_b^\dagger{\cal P}) \rangle_{\text{FS}} ,
\end{equation}
where ${\cal N}_0$ is the density of states at the Fermi energy, and
${\cal P}$ projects onto the states at the Fermi surface. 

The off-diagonal components in Eq.~(\ref{eq:det}) account for the
overlap between the on-site and extended \textit{s}-wave states.
Close to the Brillouin-zone center, the form factor of the extended
\textit{s}-wave states is 
\begin{equation}
c_x c_y + c_y c_z + c_z c_x \cong{} 3 - |\mathbf{k}|^2a^2 .
\end{equation}
Assuming weak mass anisotropy of the quadratic bands, 
the extended \textit{s}-wave potentials should
  open an approximately isotropic gap at the Fermi
  surface; the gap opened by the on-site potential is always
  isotropic.  Accordingly, the
  response of the system to the on-site and 
  the extended  
  \textit{s}-wave gaps will be very similar, and we expect the
  susceptibilities to be proportional, i.e.,
$\chi_{ee}\approx r^2\chi_{oo}$ and $\chi_{oe}\approx r\chi_{oo}$,
where $r$ is the ratio of the gap opened by the extended to the
on-site potential. The determinantal equation then reduces to 
\begin{equation}
\chi_{oo} = \frac{1}{g_o + r^2g_e} .
\end{equation}
For sufficiently large $U_0$, the on-site repulsion will dominate
over the attractive extended \text{s}-wave pairing in Eqs.\ (\ref{eq:exts.A1g})--(\ref{eq:exts.T2g}), and
the effective coupling constant $g_o+r^2g_e$ will be repulsive. As such,
we do not expect pairing in the
$A_{1g}$ or $T_{2g}$ channels in the strong-coupling limit.

\subsection{$E_g$ channels} \label{sec:Eg}

We now turn our attention to the extended \textit{s}-wave $E_g$ state. Since the $E_g$ pairing avoids the on-site Hubbard repulsion $U_0$, the interaction potential in this channel is always attractive, and it should be favored for sufficiently large $U_0$. In the following, we consider which $E_g$ pairing state is expected to be realized. The $E_{g}$ pairing channel being two dimensional, the
properties of the superconducting state are determined by a
two-component order parameter
${\Delta}_{E_{g}}\equiv(\Delta_1,\Delta_2)$. A general Landau
free-energy expansion in terms of these parameters suggests three
possible ground states: $(1,0)$, $(0,1)$, and $(1,i)$
\cite{BWW16,SiU91}. The free energies of the $(1,0)$ and $(0,1)$ states are
not expected to be the same as the two states are not related by any
point-group operation \cite{BAM18,RGF19}. The third state breaks TRS due to the imaginary number $i$
and thus has BFSs beyond infinitesimal coupling strength \cite{ABT17,BAM18}.

Within the BCS formalism, the mean-field-decoupled pairing interaction in the $E_{g}$ channel takes the form
\begin{align}
&H_{\mathrm{pair}}^{\mathrm{BCS}}
  = \frac{1}{2 N} \sum_{\mathbf{k},\mathbf{k}^\prime} \sum_{m=1}^2
    \bigg[ \Delta_{m}(\mathbf{k}) f(\mathbf{k}^\prime)\, c_{\mathbf{k}^\prime}^\dagger
    \Bar{E}_{m} c_{-\mathbf{k}^\prime}^{\dagger T} \nonumber \\
&\quad{}+ \Delta_{m}^*(\mathbf{k^\prime}) f(\mathbf{k})\, c_{-\mathbf{k}}^{ T} \Bar{E}_{m}^{\dagger} c_{\mathbf{k}}
  + \frac{\Delta_{m}(\mathbf{k})\Delta_{m}^*(\mathbf{k^\prime})}{ V_{0}} \bigg] ,
\end{align}
with the two components of the two-dimensional $E_g$ order parameter
\begin{equation}
  \Delta_{1,2}(\mathbf{k})= -V_{0} f(\mathbf{k})\, \langle c_{-\mathbf{k}}^T\Bar{E}_{1,2}^{\dagger} c_{\mathbf{k}}\rangle .
\end{equation}
Here, $f(\mathbf{k}) = c_x c_y + c_y c_z + c_z c_x$ is the
extended \textit{s}-wave form factor, $V_0$ is the absolute value of the interaction
strength given by Eq.~(\ref{eq:exts.Eg}), and 
$\Bar{E}_{1,2}$ are the pairing matrices in the $E_{g}$ channel of $j=3/2$ fermions, see Table~\ref{tab:os} and Eqs.\ (\ref{eq:J1})--(\ref{eq:J3}) in Appendix~\ref{sec:derivation_hlk}.

To study superconductivity, we numerically solve the gap equation at $T=0$,
\begin{equation}
  \Delta_{m} = \frac{V_{0}}{2 N} \sum_{\mathbf{k},i\,\in\, \mathrm{occ}}
    \frac{\partial|E_{\mathbf{k},i}|}{\partial \Delta_{m}} ,
\label{eq:BCSgapeq}
\end{equation}
where $\Delta_{m} = N^{-1} \sum_{\mathbf{k}}\Delta_{m}(\mathbf{k})$,
$m = 1,2$, $i$ represents the band index, and the sum is over all occupied states, i.e., all states with $E_{\mathbf{k},i}<0$. The derivatives can be calculated in analytical form since the problem of finding the quasiparticle energies $E_{\mathbf{k},i}$ reduces to the solution of a quartic equation. Details of the numerical method are relegated to Appendix~\ref{sec:BCS}.

\begin{figure*}
\includegraphics[width=0.95\textwidth]{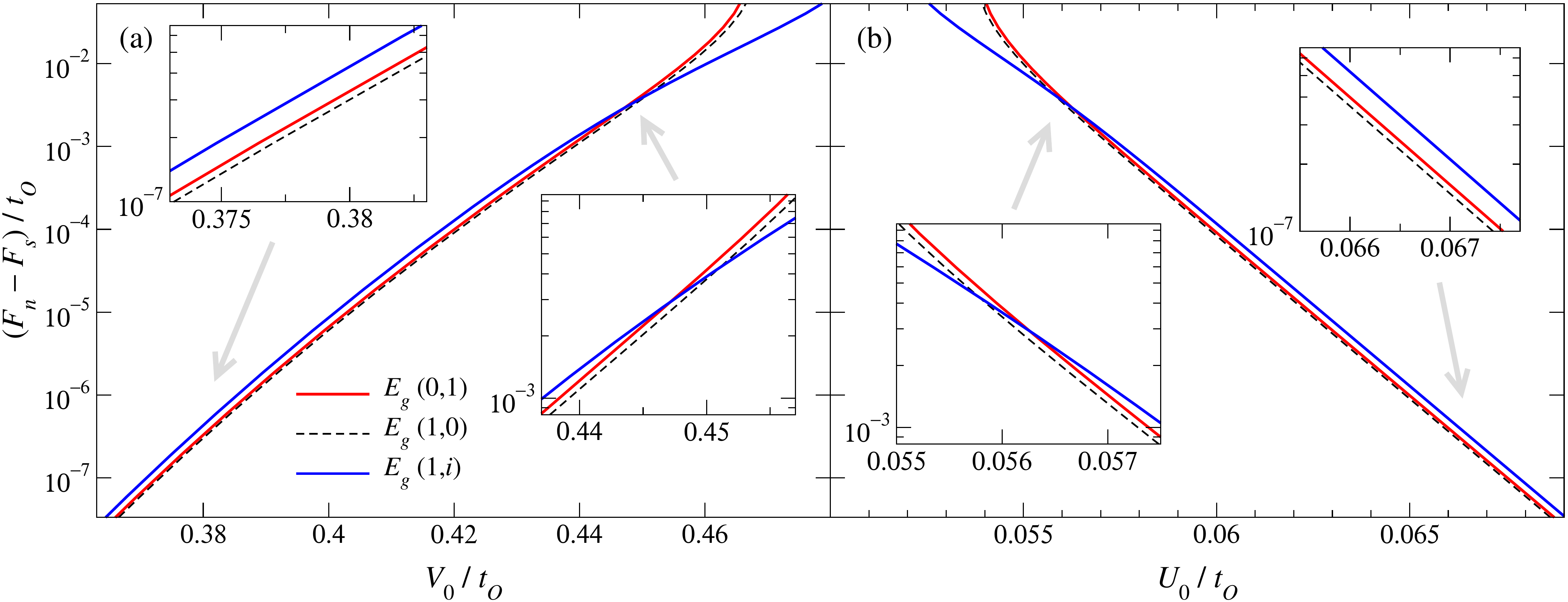}
\caption{\label{fig.BCS1} Condensation energy $F_{n}-F_{s}$ per
    unit cell for the three $E_{g}$ pairing states
as a function of (a) the interaction strength $V_{0}$ and (b) the Hubbard repulsion $U_0$. The insets show close-ups of regimes with weak and strong interaction. Note that the energetically preferred state corresponds to the \emph{largest} value.
Here, parameters $t_{1}=0.321$, $t_{2}=-0.021$, $t_{1}^\prime=-0.013$,
$t_{2}^\prime=0.004$, $t_{3}^\prime=0.008$ are used in the numerical
calculation. This corresponds to the band structure presented in
Fig.~\ref{fig:SM_LK}.}
\end{figure*}

The free energy per unit cell at $T=0$, i.e., the internal
energy per unit cell, reads as 
\begin{equation}
F=- \frac{1}{N}\sum_{\mathbf{k},i\,\in\, \mathrm{occ}}|E_{\mathbf{k},i}|+ \sum_{m=1}^{2}\frac{|\Delta_{m}|^2}{V_{0}}.
\label{eq:BCSfreeen}
\end{equation}
We compare the free energies for the three pairing states and plot the
free-energy gain, i.e., the condensation energy, on a logarithmic scale as a function of $V_{0}$ and
of $U_0$ in Fig.\ \ref{fig.BCS1}. For weak interactions $V_{0}$ and thus
small gap, the energy gain is maximal for the TRS-broken $(1,i)$
state. Increasing $V_{0}$, a first-order transition occurs to the
TRS-preserving $(0,1)$ state.

We can understand this result as follows: from Sigrist and Ueda
\cite{SiU91}, $(1,i)$ is expected to be the most stable state in the weak-coupling limit since it has point nodes and thus lower density of states close to the Fermi energy than the $(1,0)$ and $(0,1)$ states with line nodes.
For strong pairing interactions, however, the $(1,i)$
state develops large BFSs, which lead to large density of states (DOS)
and is thus no longer expected to be favored. The TRS-preserving
$(0,1)$ state is found to be more stable than the also TRS-preserving
$(1,0)$ state. They both have two line nodes but for the $(1,0)$ state
these nodes cross each other, whereas for $(0,1)$ they do not. The
crossing leads to higher DOS at the Fermi energy and is thus
disfavored~\cite{SiU91}. 

In Fig.\ \ref{fig.BCS1}(a), the data for small $V_{0}$ also show the expected weak-coupling behavior $F_{n}-F_{s} \sim e^{-A/V_{0}}$ at $T=0$ with some constant $A$; see also Appendix \ref{sec:BCS}. It is thus safe to extrapolate this curve down to zero interaction, which is not done here, though. In Fig.\ \ref{fig.BCS1}(b), the energy gain vs.\ the Hubbard repulsion $U_0$ shows nearly linear behavior, which follows from the fact that $\log (F_{n}-F_{s})$ is linear in $1/V_{0}$ in weak-coupling BCS theory and that $V_{0}$ is inversely proportional to $U_0$; see Eq.~(\ref{eq:exts.Eg}).
The energies in Fig.\ \ref{fig.BCS1} are given in units of $t_O$. To
estimate the absolute energy scale, we note that the band width of the four
bands in the model is roughly $2.5\,t_O$. Recent band-structure calculations for various
pyrochlore iridates by Antonov \textit{et al.}\ \cite{ABK20} predict
band widths of about $600\,\mathrm{meV}$ to $800\,\mathrm{meV}$. This
yields $t_O \approx 300\,\mathrm{meV}$. Using this value, we
find that the condensation energy in Fig.\ \ref{fig.BCS1} is
comparable to that predicted by weak-coupling BCS theory in
elemental superconductors.

\begin{figure}
\includegraphics[width=0.95\columnwidth]{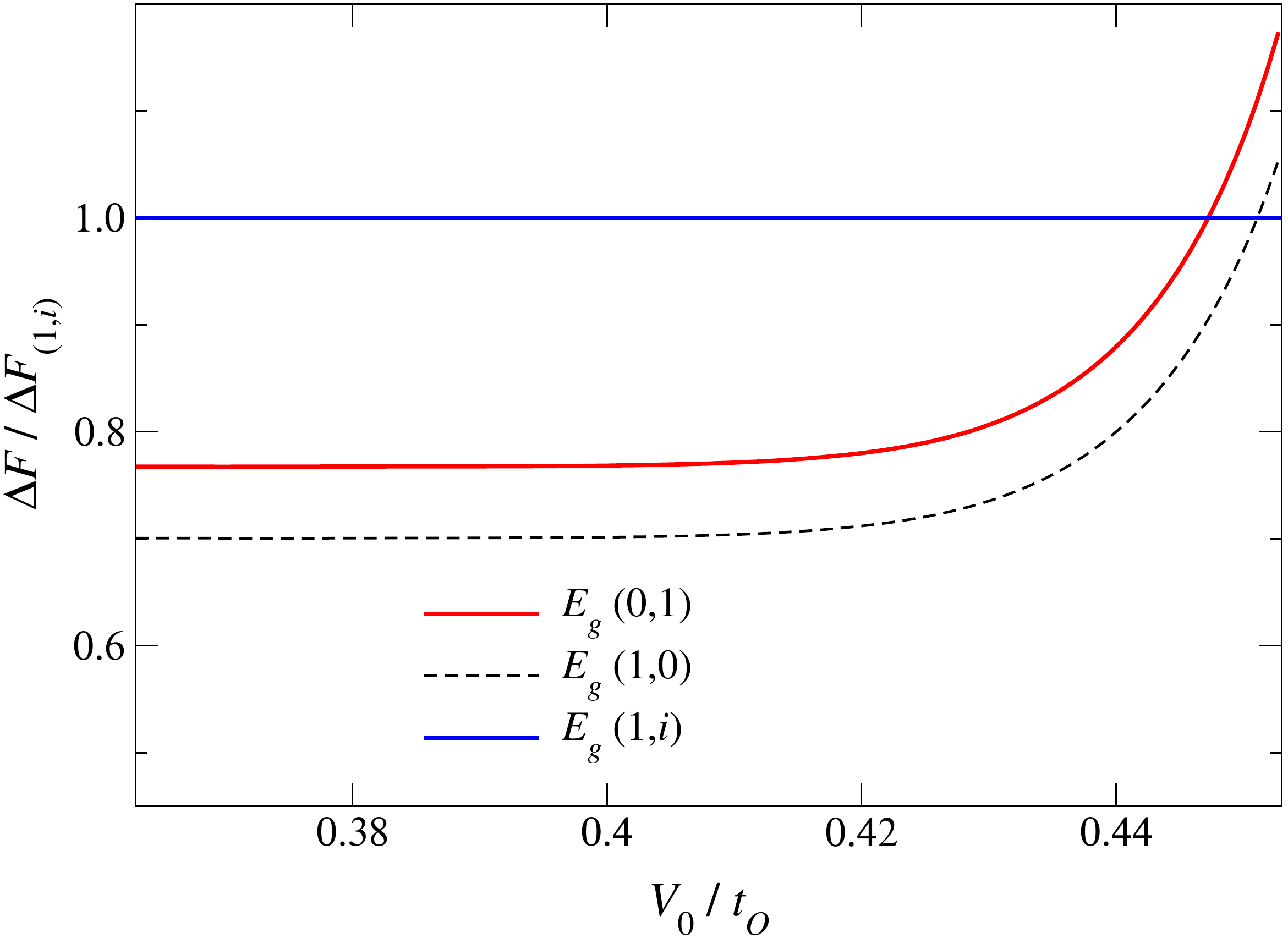}
\caption{\label{fig.BCS2} Ratios of the condensation energies $\Delta F \equiv F_n-F_s$ for the $E_g$ pairing states $(0,1)$, $(1,0)$, and $(1,i)$ to the condensation energy $\Delta F_{(1,i)}$ for the $(1,i)$ state. The parameters are the same as in Fig.~\ref{fig.BCS1}.}
\end{figure}

The differences in condensation energy of the various pairing
states in Fig.\ \ref{fig.BCS1} look rather small. This is in fact a misleading impression of the logarithmic plot. The relevant energy scale is the condensation energy itself. In Fig.\ \ref{fig.BCS2}, we therefore plot the ratio of $\Delta F \equiv F_n - F_s$ for the $(0,1)$, $(1,0)$, and $(1,i)$ states to $\Delta F$ for the $(1,i)$ state, which is favored over much of the considered range of $V_0$. Evidently, the energetic separation between the three states is sizable on the relevant energy scale.

\section{Conclusions}

In this work we have proposed the Hubbard model on the
pyrochlore lattice as a minimal tight-binding model in which to
study the superconductivity of emergent $j=3/2$ fermions. In
particular, we have demonstrated that doping the strong-coupling
limit of the half-filled Hubbard model on the pyrochlore lattice
generates an attractive interaction in the extended \textit{s}-wave quintet
$E_g$ channel. This attractive interaction results solely from the
Hubbard repulsion. The main point here is that pairing in the $E_g$ channel
avoids a local repulsive interaction and so is driven by non-local attractive
magnetic interactions.
For sufficiently strong on-site interaction, the $E_g$ pairing channel will be
favored over competing states in the $A_{1g}$ and $T_{2g}$
channels. Our numerical calculation shows that this $E_g$ pairing
state likely breaks time-reversal symmetry, and hence will support BFSs.
The time-reversal-symmetry-breaking state is compatible with the $d+id$ state found for a quasi-two-dimensional model~\cite{Laurell2017}.

Our analysis has focused entirely on pairing in the low-energy
  $j=3/2$ states, which emerge from the characteristic
  tetrahedral structural elements
  of the pyrochlore lattice. However, close to the boundaries of the $j=3/2$ semimetal phase shown
in Fig.~\ref{fig:SM_LK}(b), doping the pyrochlore
lattice will typically produce Fermi pockets of other bands elsewhere
in the Brillouin zone. Since these states do not generally have
  $j=3/2$ character,
care must be taken in considering the significance of the pairing
interaction in these regions.
In particular, the condition that the gap be nonzero at the zone
  center is less relevant, and the restriction to \textit{s}-wave-like states
  is no longer justified.
 Hence, it is promising to search for metallic pyrochlores with small Fermi pockets around the $\Gamma$ point.

\acknowledgments
S. K. was supported by JSPS KAKENHI Grant No.\ JP19K14612 and by the CREST project (JPMJCR16F2, JPMJCR19T2) from Japan Science and Technology Agency (JST).
A. B. and C. T. gratefully acknowledge financial support by the
Deut\-sche For\-schungs\-ge\-mein\-schaft through the Collaborative
Research Center SFB 1143, Project A04, and the Cluster of Excellence
on Complexity and Topology in Quantum Matter ct.qmat (EXC~2147).
P. M. R. B. is grateful for the hospitality of Nagoya University,
where part of this work was performed. P. M. R. B. was supported by
the Marsden Fund Council from Government funding, managed by Royal
Society Te Ap\=arangi. 

\appendix

\begin{figure}
\includegraphics[width=0.20\textwidth]{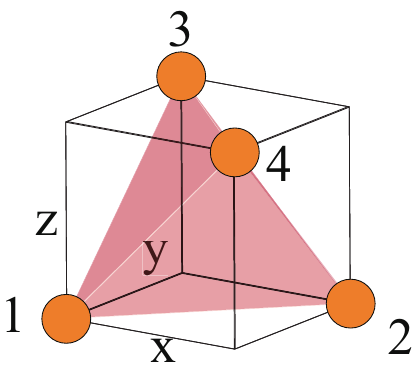}
\caption{Positions of sublattice $a=1,2,3,4$ in the elementary tetrahedron.}
\label{fig:bond}
\end{figure}

\section{Derivation of Luttinger-Kohn Hamiltonian}
\label{sec:derivation_hlk}

In this appendix, we review the derivation of the Luttinger-Kohn Hamiltonian from the pyrochlore lattice~\cite{Goswami2017}. The Hamiltonian is obtained by projecting out the $j=1/2$ subspaces and then expanding up to the quadratic order in momentum. To this end, we first provide the momentum-space form of the Hamiltonian $H=\sum_{a,a'} \sum_{\sigma \sigma'}\sum_\mathbf{k} c_{a,\sigma,\mathbf{k}}^{\dagger} [\hat{H}_0 (\mathbf{k}) + \hat{H}_{\rm SO}(\mathbf{k})]_{a, \sigma;a', \sigma'} c_{a',\sigma',\mathbf{k}}$ with
\begin{align}
\hat{H}_0(\mathbf{k}) &= F_{xy}^+ \sigma_0 \otimes \hat{\lambda}_1 +F_{yz}^+ \sigma_0 \otimes \hat{\lambda}_4
  + F_{zx}^+ \sigma_0 \otimes \hat{\lambda}_9  \notag \\
&\quad{} + F_{zx}^- \sigma_0 \otimes \hat{\lambda}_6+F_{yz}^- \sigma_0 \otimes \hat{\lambda}_{11} +F_{xy}^- \sigma_0 \otimes \hat{\lambda}_{13} \label{eq:pyrochlore_h0}
\end{align}
and
\begin{align}
\hat{H}_{\rm SO}(\mathbf{k})
  &= (G_{xy}^+ \sigma_x -G_{xy}^+ \sigma_y + K_{xy}^- \sigma_z ) \otimes \hat{\lambda}_2 \notag \\
&\quad{}+ (G_{yz}^+ \sigma_y -G_{yz}^+ \sigma_z + K_{yz}^- \sigma_x ) \otimes \hat{\lambda}_5 \notag \\
&\quad{}+ (G_{zx}^+ \sigma_z -G_{zx}^+ \sigma_x + K_{zx}^- \sigma_y ) \otimes \hat{\lambda}_{10} \notag \\
&\quad{}+ (G_{zx}^- \sigma_z +G_{zx}^- \sigma_x + K_{zx}^+ \sigma_y ) \otimes \hat{\lambda}_7 \notag \\
&\quad{}- (G_{yz}^- \sigma_y +G_{yz}^- \sigma_z + K_{yz}^+ \sigma_x ) \otimes \hat{\lambda}_{12} \notag \\
&\quad{}+ (G_{xy}^- \sigma_x +G_{xy}^- \sigma_y + K_{xy}^+ \sigma_z ) \otimes \hat{\lambda}_{14}, \label{eq:pyrochlore_hsoc}
\end{align}
where $c_{a,\sigma,\mathbf{k}}$ is a fermion annihilation operator for sublattice $a=1,2,3,4$, see Fig.~\ref{fig:bond}, and spin $\sigma = {\uparrow}, \downarrow$. The momentum dependence is represented by functions $F_{ij}^{\pm}$, $G_{ij}^{\pm}$, and $K_{ij}^{\pm}$ such that
\begin{align}
F_{ij}^{\pm} &= 2 t_1 \cos(k_i \pm k_j) + 4 t_1' \cos(2 k_l) \cos(k_i \mp k_j), \\ 
G_{ij}^{\pm} &= -2 t_2 \cos(k_i \pm k_j) \nonumber \\
&\quad{}+ 4 (t_2'+t_3') \cos(2 k_l) \cos(k_i \mp k_j), \\
K_{ij}^{\pm} &= 4 (t_2'-t_3') \cos(2 k_l) \sin(k_i \mp k_j), 
\end{align}
where $l \in \{x,y,z\} \setminus \{i,j\}$ and the lattice constant has been set to $a=4$.
$\sigma_0$ is the $2 \times 2$ identity matrix, $\sigma_i$ are the Pauli matrices, and $\hat{\lambda}_j$ are the $\mathrm{SU}(4)$ generators 
\begin{align}
\hat{\lambda}_1 &= \begin{pmatrix} 0 & 1 & 0 & 0 \\ 1 & 0&0&0 \\ 0 & 0&0&0 \\ 0 & 0&0&0 \end{pmatrix},
&\hat{\lambda}_2 &= \begin{pmatrix} 0 & -i & 0 & 0 \\ i & 0&0&0 \\ 0 & 0&0&0 \\ 0 & 0&0&0 \end{pmatrix}, \\
\hat{\lambda}_3 &= \begin{pmatrix} 1 & 0 & 0 & 0 \\ 0 & -1&0&0 \\ 0 & 0&0&0 \\ 0 & 0&0&0 \end{pmatrix}, &\hat{\lambda}_4 &= \begin{pmatrix} 0 & 0 & 1 & 0 \\ 0 & 0&0&0 \\ 1 & 0&0&0 \\ 0 & 0&0&0 \end{pmatrix}, \\
\hat{\lambda}_5 &= \begin{pmatrix} 0 & 0 & -i & 0 \\ 0 & 0&0&0 \\ i & 0&0&0 \\ 0 & 0&0&0 \end{pmatrix}, &\hat{\lambda}_6 &= \begin{pmatrix} 0 & 0 & 0 & 0 \\ 0 & 0&1&0 \\ 0 & 1&0&0 \\ 0 & 0&0&0 \end{pmatrix}, \\
\hat{\lambda}_7 &= \begin{pmatrix} 0 & 0 & 0 & 0 \\ 0 & 0&-i&0 \\ 0 & i&0&0 \\ 0 & 0&0&0 \end{pmatrix}, &\hat{\lambda}_8 &= \frac{1}{\sqrt{3}}\begin{pmatrix} 1 & 0 & 0 & 0 \\ 0 & 1&0&0 \\ 0 & 0&-2&0 \\ 0 & 0&0&0 \end{pmatrix}, \\
\hat{\lambda}_9 &= \begin{pmatrix} 0 & 0 & 0 & 1 \\ 0 & 0&0&0 \\ 0 & 0&0&0 \\ 1 & 0&0&0 \end{pmatrix}, &\hat{\lambda}_{10} &= \begin{pmatrix} 0 & 0 & 0 & -i \\ 0 & 0&0&0 \\ 0 & 0&0&0 \\ i & 0&0&0 \end{pmatrix}, \\
\hat{\lambda}_{11} &= \begin{pmatrix} 0 & 0 & 0 & 0 \\ 0 & 0&0&1 \\ 0 & 0&0&0 \\ 0 & 1&0&0 \end{pmatrix}, &\hat{\lambda}_{12} &= \begin{pmatrix} 0 & 0 & 0 & 0 \\ 0 & 0&0&-i \\ 0 & 0&0&0 \\ 0 & i &0&0 \end{pmatrix}, \\
\hat{\lambda}_{13} &= \begin{pmatrix} 0 & 0 & 0 & 0 \\ 0 & 0&0&0 \\ 0 & 0&0&1 \\ 0 & 0&1&0 \end{pmatrix},
&\hat{\lambda}_{14} &= \begin{pmatrix} 0 & 0 & 0 & 0 \\ 0 & 0&0&0 \\ 0 & 0&0&-i \\ 0 & 0 &i&0 \end{pmatrix}, \\
\hat{\lambda}_{15} &= \frac{1}{\sqrt{6}}\begin{pmatrix} 1 & 0 & 0 & 0 \\ 0 & 1&0&0 \\ 0 & 0&1&0 \\ 0 & 0 &0&-3 \end{pmatrix} . \hspace{-4em}
\end{align}
We first examine the band degeneracy at the $\Gamma$ point. In the absence of spin-orbit coupling ($t_2=t_2'=t_3'=0$), the eight bands split into sixfold and twofold degenerate bands. This can be seen by diagonalizing $\mathcal{H}_0(\mathbf{k})$ through the unitary transformation
\begin{equation}
S_1^{\dagger} \hat{H}_0({\bf 0}) S_1=-2(t_1 + 2t_1') \sqrt{6} \; \sigma_0 \otimes \hat{\lambda}_{15},
\end{equation}
with
\begin{equation}
S_1 = \frac{1}{2}\, \sigma_0 \otimes
  \begin{pmatrix} 1&-1&1&-1 \\ -1&1&1&-1 \\ 1&1&-1&-1 \\ -1&-1&-1&-1 \end{pmatrix} .
\end{equation}
When the spin-orbit couplings are turned on, the sixfold degeneracy further splits into twofold and fourfold degenerate bands:
\begin{align}
S_2^{\dagger} &S_1^{\dagger} [\mathcal{H}_0({\bf 0})+\mathcal{H}_{\rm SO} ({\bf 0})] S_1 S_2 \notag \\
&= -2(t_1 + 2t_1') \sqrt{6} \; \sigma_0 \otimes \hat{\lambda}_{15} \notag \\
&\quad{}- 4[t_2-2(t_2'+t_3')] \sqrt{3} \; \sigma_0 \otimes \hat{\lambda}_{8},
\end{align}
under the unitary transformation with
\begin{equation}
  S_2 = \begin{pmatrix} 
  \frac{i}{2} & \frac{i}{2} & \frac{1}{2\sqrt{3}} & -\frac{1}{2 \sqrt{3}}& 0 &-\frac{1}{\sqrt{3}}& 0& 0 \\
    -\frac{1}{2\sqrt{3}}& \frac{1}{2\sqrt{3}}&  -\frac{i}{2}& -\frac{i}{2}& \frac{1}{\sqrt{3}} & 0 & 0 & 0 \\
     \frac{1}{2} & \frac{1}{2}& \frac{i}{2\sqrt{3}}& -\frac{i}{2 \sqrt{3}}& 0& -\frac{i}{\sqrt{3}}& 0& 0 \\
     \frac{i}{2 \sqrt{3}}& -\frac{i}{2\sqrt{3}} & \frac{1}{2}& \frac{1}{2}& -\frac{i}{\sqrt{3}}& 0& 0& 0 \\
      \frac{1}{\sqrt{3}}& -\frac{1}{\sqrt{3}} & 0& 0& \frac{1}{\sqrt{3}}& 0& 0& 0 \\
       0& 0& \frac{1}{\sqrt{3}}& -\frac{1}{\sqrt{3}}& 0& \frac{1}{\sqrt{3}}& 0& 0 \\
      0& 0& 0& 0& 0& 0& 1& 0 \\
       0& 0& 0& 0& 0& 0& 0 & 1 
       \end{pmatrix}.
\end{equation}
As a result, the eight energy bands are split into a $j=3/2$ quartet and two $j=1/2$ doublets. Since we are interested in the $j=3/2$ quartet, we hereafter project the Hamiltonian onto the $j=3/2$ subspace and discard the $j=1/2$ doublets. The dispersion relation around the $\Gamma$ point can be obtained by expanding the projected Hamiltonian up to the quadratic order in momentum.
 Applying yet another unitary transformation with
\begin{equation}
S_3 = \frac{1}{2}\begin{pmatrix} 1-i & -1-i & 0 & 0 \\ 1-i & 1+i & 0 & 0 \\ 0 & 0 & -1-i & 1-i  \\ 0 & 0 & 1+i & 1-i  \end{pmatrix}
\end{equation}
 to the projected Hamiltonian results in a Luttinger-Kohn Hamiltonian of the canonical form
\begin{align}
\hat{H}_{\rm LK} (\mathbf{k}) &= \left( E_0+ \alpha' |\mathbf{k}|^2 \right) \mathbbm{1}_{4 \times 4} \notag \\
&\quad{} + \beta' \left(\sqrt{3}k_y k_z \hat{\Gamma}_1 +\sqrt{3} k_z k_x \hat{\Gamma}_2 +\sqrt{3} k_x k_y \hat{\Gamma}_3 \right) \notag \\
&\quad{} + \gamma' \left[ \frac{\sqrt{3}(k_x^2 -k_y^2)}{2}\, \hat{\Gamma}_4
  + \frac{(2k_z^2 -k_x^2 - k_y^2)}{2}\, \hat{\Gamma}_5\right] , \label{eq:hlk_appendix}
\end{align}
with
\begin{align}
E_0 &= -2 \left[ t_1+2(t_2+t_1')-4(t_2'+t_3') \right] , \\
\alpha' &= \frac{2}{3}\, [ t_1 +2t_2 +6t_1'-12(t_2'+t_3') ], \\
\beta' &=  -\frac{2}{3} \left[ t_1+2(t_2-t_1'-2t_2'+6t_3') \right] , \\
\gamma' &= -\frac{2}{3}\, [ t_1-t_2 -6 (t_1'+t_2'+t_3') ] .
\end{align}
Five mutually anticommuting $\Gamma$ matrices are defined as
\begin{align}
\hat{\Gamma}_1 &= \frac{1}{\sqrt{3}}\, \{\hat{J}_y, \hat{J}_z \} ,
\hat{\Gamma}_2 = \frac{1}{\sqrt{3}}\, \{\hat{J}_z, \hat{J}_x \},
\hat{\Gamma}_3 = \frac{1}{\sqrt{3}}\, \{\hat{J}_x, \hat{J}_y \}, \\ 
\hat{\Gamma}_4 &= \frac{\hat{J}_x^2 -  \hat{J}_y^2}{\sqrt{3}},
\hat{\Gamma}_5 =\frac{2\hat{J}_z^2-\hat{J}_x^2 -  \hat{J}_y^2}{3},
\end{align}
where the $\hat{J}_\mu$ are the spin-$3/2$ matrices
\begin{align}
\hat{J}_x &= \frac{1}{2}\begin{pmatrix} 0 & \sqrt{3} & 0 & 0 \\ \sqrt{3} & 0 & 2 & 0 \\0 & 2 & 0 & \sqrt{3} \\ 0 & 0 & \sqrt{3} & 0\end{pmatrix} ,
\label{eq:J1} \\ 
\hat{J}_y &= \frac{i}{2}\begin{pmatrix} 0 & -\sqrt{3} & 0 & 0 \\ \sqrt{3} & 0 & -2 & 0 \\0 & 2 & 0 & -\sqrt{3} \\ 0 & 0 & \sqrt{3} & 0\end{pmatrix}, \\
\hat{J}_z &= \frac{1}{2}\begin{pmatrix} 3 & 0 & 0 & 0 \\ 0 & 1 & 0 & 0 \\0 & 0 & -1 & 0 \\ 0 & 0 & 0 & -3\end{pmatrix}.
\label{eq:J3}
\end{align}
Comparing Eq.~(\ref{eq:hlk_appendix}) with Eq.~(\ref{eq:HLK}), we obtain the relationship between the coefficients in the two Hamiltonians,
\begin{equation}
 \alpha' = \left(\alpha + \frac{5}{4}\, \beta \right), \quad
 \beta' = \gamma, \quad
 \gamma' = \beta. 
\end{equation}
From the elements of the full unitary transformation matrix $S_1S_2S_3$, we can express the projection of the annihilation operators in the 
site-spin basis onto the low-energy $j=3/2$ subspace at the Brillouin
zone center:
\begin{align}
c_{{\bf 0},1,\uparrow} &= \frac{i}{2}\, c_{{\bf 0},{3}/{2}} - \frac{1+i}{2 \sqrt{3}}\, c_{{\bf 0},{1}/{2}}
  - \frac{1}{2\sqrt{3}}\, c_{{\bf 0},-{1}/{2}},
\label{eq:coefficients1} \\
c_{{\bf 0},1,\downarrow} &= \frac{i}{2\sqrt{3}}\, c_{{\bf 0},{1}/{2}} - \frac{1+i}{2 \sqrt{3}}\, c_{{\bf 0},-{1}/{2}}
  - \frac{1}{2}\, c_{{\bf 0},-{3}/{2}}, \\
c_{{\bf 0},2,\uparrow} &= -\frac{i}{2}\, c_{{\bf 0},{3}/{2}} - \frac{1+i}{2 \sqrt{3}}\, c_{{\bf 0},{1}/{2}}
  + \frac{1}{2\sqrt{3}}\, c_{{\bf 0},-{1}/{2}}, \\
c_{{\bf 0},2,\downarrow} &= -\frac{i}{2\sqrt{3}}\, c_{{\bf 0},{1}/{2}} - \frac{1+i}{2 \sqrt{3}}\, c_{{\bf 0},-{1}/{2}}
  + \frac{1}{2}\, c_{{\bf 0},-{3}/{2}}, \\
c_{{\bf 0},3,\uparrow} &= \frac{1}{2}\, c_{{\bf 0},{3}/{2}} + \frac{1+i}{2 \sqrt{3}}\, c_{{\bf 0},{1}/{2}}
  - \frac{i}{2\sqrt{3}}\, c_{{\bf 0},-{1}/{2}}, \\
c_{{\bf 0},3,\downarrow} &= \frac{1}{2\sqrt{3}}\, c_{{\bf 0},{1}/{2}} + \frac{1+i}{2 \sqrt{3}}\, c_{{\bf 0},-{1}/{2}}
  - \frac{i}{2}\, c_{{\bf 0},-{3}/{2}}, \\
c_{{\bf 0},4,\uparrow} &= -\frac{1}{2}\, c_{{\bf 0},{3}/{2}} + \frac{1+i}{2 \sqrt{3}}\, c_{{\bf 0},{1}/{2}}
  + \frac{i}{2\sqrt{3}}\, c_{{\bf 0},-{1}/{2}}, \\
c_{{\bf 0},4,\downarrow} &= -\frac{1}{2\sqrt{3}}\, c_{{\bf 0},{1}/{2}} + \frac{1+i}{2 \sqrt{3}}\, c_{{\bf 0},-{1}/{2}}
  + \frac{i}{2}\, c_{{\bf 0},-{3}/{2}},
\label{eq:coefficients8}
\end{align}
where $c_{{\bf k},n,\sigma}$ annihilates an electron with momentum ${\bf
k}$ and spin $\sigma$ at site $n$ of the tetrahedron, and $c_{{\bf
k},\tilde{s}}$ annihilates an electron with momentum ${\bf k}$ and
$\tilde{s}=-3/2,-1/2,1/2,3/2$. Although the coefficients
in Eqs.~(\ref{eq:coefficients1})--(\ref{eq:coefficients8}) will be
momentum-dependent away from ${\bf k}={\bf 0}$, we continue to use the
${\bf k}={\bf 0}$ coefficients since the $j=3/2$ description is only valid
sufficiently close to the Brillouin-zone center, where the zero-order contributions to these coefficients dominate.
Within this approximation, it follows that the coefficients $u_{a,\sigma,j}$
in Eq.~(\ref{eq:transform}) are identical to the coefficients
appearing in Eqs.~(\ref{eq:coefficients1})--(\ref{eq:coefficients8}).

\section{Interactions projected onto the $j=3/2$ subspace}
\label{sec:interaction_hlk}

In this appendix, we derive the explicit form of the on-site and nearest-neighbor interactions $V$ and $V_{a,a'}$ projected onto the $j=3/2$ subspace. These are obtained by substituting Eqs.~(\ref{eq:coefficients1})--(\ref{eq:coefficients8}) into Eqs.~(\ref{eq:Vonsite}) and (\ref{eq:Vnn}). To obtain a compact description, we employ a symmetric form presented in Ref.~\onlinecite{Boettcher2017}. A local interaction term can be written as
\begin{equation}
g_{NM} (\psi^{\dagger} N \psi)(\phi^{\dagger} M \phi),
\end{equation}
with coupling $g_{NM}$, field operators in a basis of $j=3/2$ fermions $\psi$ and $\phi$,
and $4 \times 4$ Hermitian matrices $N$ and $M$. In order to cover all possible interactions, we introduce a basis of sixteen matrices that are irreducible tensor operators of the point group $O_h$~\cite{Boettcher2017,TiB21}:
\begin{align}
\hat{E}_1 &= \hat{\Gamma}_4 , \\
\hat{E}_2 &= \hat{\Gamma}_5 , \\
\hat{T}_i &= \hat{\Gamma}_i , \\
\hat{\mathcal{J}}_i &= \frac{2}{\sqrt{5}}\, \hat{J}_i, \\
\hat{W}_i &= \frac{2 \sqrt{5}}{3} \left( \hat{J}_i^3 - \frac{41}{20} \hat{J}_i \right),  \\
\hat{W}_i' &= \frac{1}{\sqrt{3}} \left[ \hat{J}_i, \left(\hat{J}_{i+1}^2-\hat{J}_{i+2}^2 \right) \right], \\
\hat{W}_7 &= \frac{2}{\sqrt{3}}\, (\hat{J}_x\hat{J}_y\hat{J}_z+\hat{J}_z\hat{J}_y \hat{J}_x), 
\end{align}
and the $4 \times 4$ identity matrix $\mathbbm{1}$.  Here, $i=x,y,z$ and $i+1$ and $i+2$ are understood cyclically. These sixteen matrices satisfy
\begin{equation}
\Tr(\hat{A} \hat{B}) = 4\delta_{\hat{A},\hat{B}} .
\label{eq:traceorthog}
\end{equation}
Of these matrices, $\mathbbm{1}$ belongs to the irrep $A_{1g}$, $\hat{E}_1$ and $\hat{E}_2$ belong to $E_g$, $\hat{T}_i$ belong to $T_{2g}$, $\hat{\mathcal{J}}_i$ and $\hat{W}_i$ belong to $T_{1g}$, $\hat{W}_i'$ belong to $T_{2g}$, and $\hat{W}_7$ belongs to $A_{2g}$ \cite{TiB21}.

In the following, we show an explicit form of onsite and nearest-neighbor interactions using the sixteen basis matrices. We employ a vector notation with
\begin{equation}
\vec{\hat{T}} = (\hat{T}_1,\hat{T}_2,\hat{T}_3)
\end{equation}
etc.

\subsection{On-site interaction}

We first consider the on-site interaction, which is readily calculated as
\begin{equation}
V = \frac{U_0}{8}\, \hat{\mathbbm{1}} \hattimes \hat{\mathbbm{1}}
  + \frac{U_0}{24}\, \vec{\hat{T}} \hattimes \vec{\hat{T}} ,
\label{eq:onsite_explicit}
\end{equation}
where the product $\hattimes$ is defined by
\begin{align}
&\sum_{\alpha ,\alpha',\beta ,\beta'} (\hat{A} \hattimes \hat{B})_{\alpha\alpha';\beta\beta'}\,
  \psi^{\dagger}_{\alpha} \psi_{\alpha'} \psi^{\dagger}_{\beta} \psi_{\beta'}  \nonumber \\
&\quad= \sum_{\alpha ,\alpha',\beta ,\beta'} \hat{A}_{\alpha \alpha'} \hat{B}_{\beta \beta'} \,
   \psi^{\dagger}_{\alpha} \psi_{\alpha'} \psi^{\dagger}_{\beta} \psi_{\beta'} \nonumber \\
&\quad= \left(\sum_{\alpha ,\alpha'}   \psi^{\dagger}_{\alpha}\hat{A}_{\alpha \alpha'} \psi_{\alpha'} \right) \left(\sum_{\beta ,\beta'}\psi^{\dagger}_{\beta}\hat{B}_{\beta \beta'} \psi_{\beta'} \right) \nonumber \\
&\quad\equiv (\psi^{\dagger} \hat{A} \psi)(\psi^{\dagger} \hat{B} \psi)
\end{align}
 and if $\hat{A}$ and $\hat{B}$ are vectors of equal dimension, summation over their components is implied.
The results for the effective interactions in each channel reveal that the on-site Hubbard interaction is relevant for the \textit{s}-wave $A_{1g}$ and $T_{2g}$ channels, but that the \textit{s}-wave $E_g$ channels are insensitive to this interaction.

\subsection{Charge-charge interaction}
Next, we consider the nearest-neighbor interactions, given by the second term in $H_\mathrm{int}$, see Eq.\ \eqref{eq:j32int}. We make the numbers $i$, $j$ of the tetrahedra, i.e., the sites, explicit by writing
\begin{align}
&\sum_{\langle i, j \rangle_{a a'}}\sum_{\alpha,\alpha'}\sum_{\beta,\beta'}[V_{a,a'}]_{\alpha \alpha';\beta \beta'}c^\dagger_{i,\alpha}c^{}_{i,\alpha'}c^\dagger_{j,\beta}c^{}_{j,\beta'} \notag \\
&\quad= \sum_{\langle i, j \rangle_{a a'}}\sum_{\alpha,\alpha'}\sum_{\beta,\beta'}[V_{ia,ja'}]_{\alpha \alpha';\beta \beta'}c^\dagger_{i,\alpha}c^{}_{i,\alpha'}c^\dagger_{j,\beta}c^{}_{j,\beta'}
\end{align}
and symmetrize the interaction by rewriting the previous expression as
\begin{equation}
\frac{1}{2}\sum_{\langle i, j \rangle_{aa'}} \sum_{\alpha,\alpha'}\sum_{\beta,\beta'}
  [V_{ia,ja'} + V_{ja',ia}]_{\alpha \alpha';\beta \beta'}
  c^\dagger_{i,\alpha}c^{}_{i,\alpha'}c^\dagger_{j,\beta}c^{}_{j,\beta'} .
\label{eq:nnint_explicit}
\end{equation}
The interaction strength is given by Eq.\ \eqref{eq:Vnn}. It can be written in terms of expressions depending on the sites $i$, $j$ and expressions depending on the orientation of the bond $(aa')$ between the corners $a$ and $a'$ of the elementary tetrahedron, see Fig.\ \ref{fig:bond}, as
\begin{align}
&V_{ia,ja'} + V_{ja',ia} = V_{ij}^0 + \widehat{yz}^2_{aa'} V_{ij}^{1} + \widehat{xz}^2_{aa'} V_{ij}^{2} + \widehat{xy}^2_{aa'} V_{ij}^{3} \notag \\
&\qquad{}+ \widehat{yz}_{aa'} V_{ij}^{4}+ \widehat{xz}_{aa'} V_{ij}^{5}+ \widehat{xy}_{aa'} V_{ij}^{6}, \label{eq:nnintv2}
\end{align}
where
\begin{align}
 &\widehat{yz}_{aa'} = 
 \begin{cases} 
  1 & \text{if } (aa') = (13), \\
  -1 & \text{if } (aa') = (24),  \\
  0 & \text{otherwise},
 \end{cases}
  \\
 & \widehat{xz}_{aa'} = 
 \begin{cases} 
  1 & \text{if } (aa') = (14), \\
  -1 & \text{if } (aa') = (23),  \\
  0 & \text{otherwise},
 \end{cases}
  \\
 & \widehat{xy}_{aa'} = 
 \begin{cases} 
  1 & \text{if } (aa') = (12), \\
  -1 & \text{if } (aa') = (34),  \\
  0 & \text{otherwise}.
 \end{cases}
\end{align}
 Here, we take $(aa')=(a'a)$. The coefficients $V_{ij}^0$ etc.\ can be obtained by substituting the coefficients from Eqs.~(\ref{eq:coefficients1})--(\ref{eq:coefficients8}) into Eq.~(\ref{eq:Vnn}).

The charge-charge interaction is given by the first term in Eq.~(\ref{eq:Vnn}).
Substituting the coefficients from Eqs.~(\ref{eq:coefficients1})--(\ref{eq:coefficients8}) into Eq.~(\ref{eq:Vnn}), this interaction is calculated as
\begin{align}
V_{ij}^{{U_1,}0} &= \frac{U_1}{8}\, \hat{\mathbbm{1}}_i \hattimes \hat{\mathbbm{1}}_j  \notag \\
&\quad{}- \frac{U_1}{24} \left( \hat{T}_{1i} \hattimes \hat{T}_{1j}
  + \hat{T}_{2i} \hattimes \hat{T}_{2j}+\hat{T}_{3i} \hattimes \hat{T}_{3j}\right), \\
V_{ij}^{{U_1,}1} &= \frac{U_1}{12}\, \hat{T}_{1i} \hattimes \hat{T}_{1j}, \\
V_{ij}^{{U_1,}2} &= \frac{U_1}{12}\, \hat{T}_{2i} \hattimes \hat{T}_{2j}, \\
V_{ij}^{{U_1,}3} &= \frac{U_1}{12}\, \hat{T}_{3i} \hattimes \hat{T}_{3j}, \\
V_{ij}^{{U_1,}4} &= -\frac{U_1}{4 \sqrt{3}}\, \{ \hat{\mathbbm{1}}, \hat{T}_1\}_{ij}
  - \frac{U_1}{12}\, \{\hat{T}_2, \hat{T}_3\}_{ij}, \\
V_{ij}^{{U_1,}5} &= -\frac{U_1}{4 \sqrt{3}}\, \{ \hat{\mathbbm{1}}, \hat{T}_2\}_{ij}
  - \frac{U_1}{12}\, \{\hat{T}_1, \hat{T}_3\}_{ij}, \\
V_{ij}^{{U_1,}6} &=-\frac{U_1}{4 \sqrt{3}}\, \{ \hat{\mathbbm{1}}, \hat{T}_3\}_{ij}
  - \frac{U_1}{12}\, \{\hat{T}_1, \hat{T}_2\}_{ij},  
\end{align} 
where we define $\{\hat{A},\hat{B}\}_{ij} \equiv \frac{1}{2} (\hat{A}_i \hattimes \hat{B}_j + \hat{B}_i \hattimes \hat{A}_j)$.

\subsection{Heisenberg interaction}

The Heisenberg interaction is given by the second term in Eq.~(\ref{eq:Vnn}). In a similar manner, the Heisenberg interaction is also represented by the irreducible spin tensors, which yields
\begin{widetext}
\begin{align}
V_{ij}^{J,0} &= \frac{5J}{288}  \left( \hat{\mathcal{J}}_{1i} \hattimes \hat{\mathcal{J}}_{1j}
  + \hat{\mathcal{J}}_{2i} \hattimes \hat{\mathcal{J}}_{2j}+\hat{\mathcal{J}}_{3i}
    \hattimes \hat{\mathcal{J}}_{3j}\right)
  - \frac{J}{96}\, \hat{W}_{7i} \hattimes \hat{W}_{7j}, \\
V_{ij}^{J,1} &= -\frac{J}{720}\, \hat{\mathcal{J}}_{1i} \hattimes \hat{\mathcal{J}}_{1j}
  - \frac{J}{80}\, \hat{W}_{1i} \hattimes \hat{W}_{1j}
  - \frac{J}{48}\, \hat{W}_{1i}' \hattimes \hat{W}_{1j}' \notag \\
&\quad{} + \frac{J}{120}\, \{ \hat{\mathcal{J}}_1, \hat{W}_1 \}_{ij}
  - \frac{J}{8\sqrt{15}} \left(\{\hat{W}_2, \hat{W}_2'\}_{ij} - \{\hat{W}_3, \hat{W}_3'\}_{ij} \right)
  + \frac{J}{24 \sqrt{15}} \left(\{\hat{\mathcal{J}}_2, \hat{W}_2'\}_{ij}
  - \{\hat{\mathcal{J}}_3, \hat{W}_3'\}_{ij} \right), \\
V_{ij}^{J,2} &= -\frac{J}{720}\, \hat{\mathcal{J}}_{2i} \hattimes \hat{\mathcal{J}}_{2j}
  + \frac{J}{80}\, \hat{W}_{2i} \hattimes \hat{W}_{2j}
  - \frac{J}{48}\, \hat{W}_{2i}' \hattimes \hat{W}_{2j}' \notag \\
&\quad{} + \frac{J}{120}\, \{ \hat{\mathcal{J}}_2, \hat{W}_2 \}_{ij}
  - \frac{J}{8\sqrt{15}} \left(\{\hat{W}_3, \hat{W}_3'\}_{ij} - \{\hat{W}_1, \hat{W}_1'\}_{ij} \right)
  + \frac{J}{24 \sqrt{15}} \left(\{\hat{\mathcal{J}}_3, \hat{W}_3'\}_{ij}
  - \{\hat{\mathcal{J}}_1, \hat{W}_1'\}_{ij} \right), \\
V_{ij}^{J,3} &= -\frac{J}{720}\, \hat{\mathcal{J}}_{3i} \hattimes \hat{\mathcal{J}}_{3j}
  - \frac{J}{80}\, \hat{W}_{3i} \hattimes \hat{W}_{3j}
  - \frac{J}{48}\, \hat{W}_{3i}' \hattimes \hat{W}_{3j}' \notag \\
&\quad{}+ \frac{J}{120}\, \{ \hat{\mathcal{J}}_3, \hat{W}_3 \}_{ij}
  - \frac{J}{8\sqrt{15}} \left(\{\hat{W}_1, \hat{W}_1'\}_{ij} - \{\hat{W}_2, \hat{W}_2'\}_{ij} \right)
  + \frac{J}{24 \sqrt{15}} \left(\{\hat{\mathcal{J}}_1, \hat{W}_1'\}_{ij}
  - \{\hat{\mathcal{J}}_2, \hat{W}_2'\}_{ij} \right), \\
V_{ij}^{J,4} &= -\frac{11J}{720}\, \{ \hat{\mathcal{J}}_2, \hat{\mathcal{J}}_3\}_{ij}
  - \frac{J}{80}\, \{\hat{W}_2, \hat{W}_3\}_{ij}
  + \frac{J}{8 \sqrt{15}}\, \{ \hat{W}_1, \hat{W}_7\}_{ij}
  + \frac{J}{48}\, \{ \hat{W}_2', \hat{W}_3' \}_{ij} 
  +\frac{J}{16\sqrt{15}} \left(\{\hat{W}_2, \hat{W}_3'\}_{ij}
    - \{\hat{W}_3, \hat{W}_2'\}_{ij} \right) \notag \\
&\quad{} - \frac{7J}{48 \sqrt{15}}\, \{\hat{\mathcal{J}}_1, \hat{W}_7\}_{ij}
  + \frac{J}{40} \left(\{\hat{\mathcal{J}}_2, \hat{W}_3\}_{ij} + \{\hat{\mathcal{J}}_3, \hat{W}_2\}_{ij} \right)
  + \frac{J}{12 \sqrt{15}} \left(\{\hat{\mathcal{J}}_2, \hat{W}_3'\}_{ij}
    - \{\hat{\mathcal{J}}_3, \hat{W}_2'\}_{ij} \right), \\
V_{ij}^{J,5} &= -\frac{11J}{720}\, \{ \hat{\mathcal{J}}_3, \hat{\mathcal{J}}_1\}_{ij}
  - \frac{J}{80}\, \{\hat{W}_3, \hat{W}_1\}_{ij}
  + \frac{J}{8 \sqrt{15}}\, \{ \hat{W}_2, \hat{W}_7\}_{ij}
  + \frac{J}{48}\, \{ \hat{W}_3', \hat{W}_1' \}_{ij}
  + \frac{J}{16\sqrt{15}} \left(\{\hat{W}_3, \hat{W}_1'\}_{ij} - \{\hat{W}_1, \hat{W}_3'\}_{ij} \right) \notag \\
&\quad{} - \frac{7J}{48 \sqrt{15}}\, \{\hat{\mathcal{J}}_2, \hat{W}_7\}_{ij}
  + \frac{J}{40}\, \left(\{\hat{\mathcal{J}}_3, \hat{W}_1\}_{ij} + \{\hat{\mathcal{J}}_1, \hat{W}_3\}_{ij} \right)
  + \frac{J}{12 \sqrt{15}} \left(\{\hat{\mathcal{J}}_3, \hat{W}_1'\}_{ij}
    - \{\hat{\mathcal{J}}_1, \hat{W}_3'\}_{ij} \right), \\
V_{ij}^{J,6} &= -\frac{11J}{720}\, \{ \hat{\mathcal{J}}_1, \hat{\mathcal{J}}_2\}_{ij}
  - \frac{J}{80}\, \{\hat{W}_1, \hat{W}_2\}_{ij}
  + \frac{J}{8 \sqrt{15}}\, \{ \hat{W}_3, \hat{W}_7\}_{ij}
  + \frac{J}{48}\, \{ \hat{W}_1', \hat{W}_2' \}_{ij}
  + \frac{J}{16\sqrt{15}} \left(\{\hat{W}_1, \hat{W}_2'\}_{ij} - \{\hat{W}_2, \hat{W}_1'\}_{ij} \right) \notag \\
&\quad{} - \frac{7J}{48 \sqrt{15}}\, \{\hat{\mathcal{J}}_3, \hat{W}_7\}_{ij}
  + \frac{J}{40} \left(\{\hat{\mathcal{J}}_1, \hat{W}_2\}_{ij} + \{\hat{\mathcal{J}}_2, \hat{W}_1\}_{ij} \right)
  + \frac{J}{12 \sqrt{15}} \left(\{\hat{\mathcal{J}}_1, \hat{W}_2'\}_{ij}
    - \{\hat{\mathcal{J}}_2, \hat{W}_1'\}_{ij} \right).
\end{align} 

\subsection{Dzyaloshinski-Moriya interaction}

The Dzyaloshinskii-Moriya interaction is given by the third term in Eq.~(\ref{eq:Vnn}), where ${\bf d}_{ij} = - {\bf d}_{ji}$ is a vector perpendicular to the bond $(ij)$ and takes the values ${\bf d}_{12} =(1,-1,0)$, ${\bf d}_{13} =(0,1,-1)$,  ${\bf d}_{14} =(-1,0,1)$, ${\bf d}_{23} =(1,0,1)$, ${\bf d}_{24} =(0,-1,-1)$, and ${\bf d}_{34} =(1,1,0)$. Here, we have set the lattice constant to $a=4$. We obtain the interaction terms
\begin{align}
V_{ij}^{D,0} &= \frac{D}{180} \left( \hat{\mathcal{J}}_{1i} \hattimes \hat{\mathcal{J}}_{1j}
  + \hat{\mathcal{J}}_{2i} \hattimes \hat{\mathcal{J}}_{2j}
  + \hat{\mathcal{J}}_{3i} \hattimes \hat{\mathcal{J}_{3j}}\right)
  - \frac{D}{80} \left( \hat{W}_{1i} \hattimes \hat{W}_{1j}
  + \hat{W}_{2i} \hattimes \hat{W}_{2j}
  + \hat{W}_{3i} \hattimes \hat{W}_{3j} \right) \notag \\
&\quad{}+ \frac{D}{48} \left( \hat{W}_{1i}' \hattimes \hat{W}_{1j}'
  + \hat{W}_{2i}' \hattimes \hat{W}_{2j}'
  + \hat{\mathcal{W}}_{3i}' \hattimes \hat{W}_{3j}'\right)
  - \frac{D}{80} \left(\{ \hat{\mathcal{J}}_1, \hat{W}_1\}_{ij} +\{ \hat{\mathcal{J}}_2, \hat{W}_2\}_{ij}
    +\{ \hat{\mathcal{J}}_3, \hat{W}_3\}_{ij} \right)
  + \frac{D}{24}\, \hat{W}_{7i} \hattimes \hat{W}_{7j}, \\
V_{ij}^{D,1} &= -\frac{7D}{360}\, \hat{\mathcal{J}}_{1i} \hattimes \hat{\mathcal{J}}_{1j}
  + \frac{D}{80}\, \hat{W}_{1i} \hattimes \hat{W}_{1j}
  - \frac{D}{48}\, \hat{W}_{1i}' \hattimes \hat{W}_{1j}'
  + \frac{13D}{240}\, \{ \hat{\mathcal{J}}_1, \hat{W}_1 \}_{ij}
  - \frac{5D}{48\sqrt{15}} \left(\{\hat{\mathcal{J}}_2, \hat{W}_2'\}_{ij}
  - \{\hat{\mathcal{J}}_3, \hat{W}_3'\}_{ij} \right), \\
V_{ij}^{D,2} &= -\frac{7D}{360}\, \hat{\mathcal{J}}_{2i} \hattimes \hat{\mathcal{J}}_{2j}
  + \frac{D}{80}\, \hat{W}_{2i} \hattimes \hat{W}_{2j}
  - \frac{D}{48}\, \hat{W}_{2i}' \hattimes \hat{W}_{2j}'
  + \frac{13D}{240}\, \{ \hat{\mathcal{J}}_2, \hat{W}_2 \}_{ij}
  - \frac{5D}{48\sqrt{15}} \left(\{\hat{\mathcal{J}}_3, \hat{W}_3'\}_{ij}
  - \{\hat{\mathcal{J}}_1, \hat{W}_1'\}_{ij} \right), \\
V_{ij}^{D,3} &= -\frac{7D}{360}\, \hat{\mathcal{J}}_{3i} \hattimes \hat{\mathcal{J}}_{3j}
  + \frac{D}{80}\, \hat{W}_{3i} \hattimes \hat{W}_{3j}
  - \frac{D}{48}\, \hat{W}_{3i}' \hattimes \hat{W}_{3j}'
  + \frac{13D}{240}\, \{ \hat{\mathcal{J}}_3, \hat{W}_3 \}_{ij}
  - \frac{5D}{48\sqrt{15}} \left(\{\hat{\mathcal{J}}_1, \hat{W}_1'\}_{ij}
    - \{\hat{\mathcal{J}}_2, \hat{W}_2'\}_{ij} \right), \\
V_{ij}^{D,4} &= \frac{D}{90}\, \{ \hat{\mathcal{J}}_2, \hat{\mathcal{J}}_3\}_{ij}
  - \frac{D}{40}\, \{\hat{W}_2, \hat{W}_3\}_{ij}
  - \frac{D}{8 \sqrt{15}}\, \{ \hat{W}_1, \hat{W}_7\}_{ij}
  - \frac{D}{6 \sqrt{15}}\, \{\hat{\mathcal{J}}_1, \hat{W}_7\}_{ij}
  + \frac{5D}{48\sqrt{15}} \left(\{\hat{\mathcal{J}}_2, \hat{W}_3'\}_{ij}
    - \{\hat{\mathcal{J}}_3, \hat{W}_2'\}_{ij} \right) \notag \\
&\quad{} - \frac{D}{24}\, \{ \hat{W}_2', \hat{W}_3' \}_{ij}
  - \frac{D}{80} \left(\{\hat{\mathcal{J}}_2, \hat{W}_3\}_{ij}
    + \{\hat{\mathcal{J}}_3, \hat{W}_2\}_{ij} \right), \\
V_{ij}^{D,5} &= \frac{D}{90}\, \{ \hat{\mathcal{J}}_3, \hat{\mathcal{J}}_1\}_{ij}
  - \frac{D}{40}\, \{\hat{W}_3, \hat{W}_1\}_{ij}
  - \frac{D}{8 \sqrt{15}}\, \{ \hat{W}_2, \hat{W}_7\}_{ij}
  - \frac{D}{6 \sqrt{15}}\,  \{\hat{\mathcal{J}}_2, \hat{W}_7\}_{ij}
  + \frac{5D}{48\sqrt{15}} \left(\{\hat{\mathcal{J}}_3, \hat{W}_1'\}_{ij}
    - \{\hat{\mathcal{J}}_1, \hat{W}_3'\}_{ij}   \right) \notag \\
&\quad{} - \frac{D}{24}\, \{ \hat{W}_3', \hat{W}_1' \}_{ij}
  - \frac{D}{80} \left(\{\hat{\mathcal{J}}_3, \hat{W}_1\}_{ij}
    + \{\hat{\mathcal{J}}_1, \hat{W}_3\}_{ij} \right),  \\
V_{ij}^{D,6} &= \frac{D}{90}\, \{ \hat{\mathcal{J}}_1, \hat{\mathcal{J}}_2\}_{ij}
  - \frac{D}{40}\, \{\hat{W}_1, \hat{W}_2\}_{ij}
  - \frac{D}{8 \sqrt{15}}\, \{ \hat{W}_3, \hat{W}_7\}_{ij}
  - \frac{D}{6 \sqrt{15}}\, \{\hat{\mathcal{J}}_3, \hat{W}_7\}_{ij}
  + \frac{5D}{48\sqrt{15}} \left(\{\hat{\mathcal{J}}_1, \hat{W}_2'\}_{ij}
    - \{\hat{\mathcal{J}}_2, \hat{W}_1'\}_{ij} \right) \notag \\
&\quad{} - \frac{D}{24}\, \{ \hat{W}_1', \hat{W}_2' \}_{ij}
  - \frac{D}{80} \left(\{\hat{\mathcal{J}}_1, \hat{W}_2\}_{ij}
    + \{\hat{\mathcal{J}}_2, \hat{W}_1\}_{ij} \right).
\end{align}

\subsection{Traceless symmetric interaction}

The traceless symmetric interaction is given by the fourth term in Eq.~(\ref{eq:Vnn}), where $\Gamma^{\mu\nu}_{ij}$ splits into diagonal and off-diagonal parts: $\Gamma^{\mu\nu}_{ij}=d^\mu_{ij} d^\nu_{ij}(\Gamma_0\delta_{\mu\nu}+\Gamma_1[1-\delta_{\mu\nu}])$. The corresponding interaction terms are given by
\begin{align}
V_{ij}^{\Gamma,0} &= \frac{13 \Gamma_0-5\Gamma_1}{720} \left( \hat{\mathcal{J}}_{1i} \hattimes \hat{\mathcal{J}}_{1j}
    + \hat{\mathcal{J}}_{2i} \hattimes \hat{\mathcal{J}}_{2j}
    + \hat{\mathcal{J}}_{3i} \hattimes \hat{\mathcal{J}}_{3j}\right)
  + \frac{\Gamma_0}{160} \left( \hat{W}_{1i} \hattimes \hat{W}_{1j}
    + \hat{W}_{2i} \hattimes \hat{W}_{2j}
    + \hat{W}_{3i} \hattimes \hat{W}_{3j}\right) \notag \\
&\quad{} + \frac{\Gamma_0}{96} \left( \hat{W}_{1i}' \hattimes \hat{W}_{1j}'
    + \hat{W}_{2i}' \hattimes \hat{W}_{2j}'
    + \hat{W}_{3i}' \hattimes \hat{W}_{3j}'\right)
  - \frac{\Gamma_0-5\Gamma_1}{240} \left(\{ \hat{\mathcal{J}}_1, \hat{W}_1\}_{ij}
    + \{ \hat{\mathcal{J}}_2, \hat{W}_2\}_{ij} 
    + \{ \hat{\mathcal{J}}_3, \hat{W}_3\}_{ij} \right) \notag \\
&\quad{}- \frac{\Gamma_0+\Gamma_1}{48}\, \hat{W}_{7i} \hattimes \hat{W}_{7j}, \\
V_{ij}^{\Gamma,1} &= -\frac{7\Gamma_0-2\Gamma_1}{360}\, \hat{\mathcal{J}}_{1i} \hattimes \hat{\mathcal{J}}_{1j}
  - \frac{3\Gamma_0+2\Gamma_1}{160}\, \hat{W}_{1i} \hattimes \hat{W}_{1j}
  - \frac{3\Gamma_0-2\Gamma_1}{96}\, \hat{W}_{1i}' \hattimes \hat{W}_{1j}'
  + \frac{\Gamma_0+\Gamma_1}{120}\, \{ \hat{\mathcal{J}}_1, \hat{W}_1 \}_{ij} \notag \\
&\quad{} - \frac{\Gamma_0}{16\sqrt{15}} \left(\{\hat{W}_2, \hat{W}_2'\}_{ij}
    - \{\hat{W}_3, \hat{W}_3'\}_{ij} \right)
  + \frac{\Gamma_0-5\Gamma_1}{48 \sqrt{15}} \left(\{\hat{\mathcal{J}}_2, \hat{W}_2'\}_{ij}
    - \{\hat{\mathcal{J}}_3, \hat{W}_3'\}_{ij} \right), \\
V_{ij}^{\Gamma,2} &= -\frac{7\Gamma_0-2\Gamma_1}{360}\, \hat{\mathcal{J}}_{2i} \hattimes \hat{\mathcal{J}}_{2j}
  - \frac{3\Gamma_0+2\Gamma_1}{160}\, \hat{W}_{2i} \hattimes \hat{W}_{2j}
  - \frac{3\Gamma_0-2\Gamma_1}{96}\, \hat{W}_{2i}' \hattimes \hat{W}_{2j}'
  + \frac{\Gamma_0+\Gamma_1}{120}\, \{ \hat{\mathcal{J}}_2, \hat{W}_2 \}_{ij} \notag \\
&\quad{} - \frac{\Gamma_0}{16\sqrt{15}} \left(\{\hat{W}_3, \hat{W}_3'\}_{ij}
    - \{\hat{W}_1, \hat{W}_1'\}_{ij} \right)
  + \frac{\Gamma_0-5\Gamma_1}{48 \sqrt{15}} \left(\{\hat{\mathcal{J}}_3, \hat{W}_3'\}_{ij}
    - \{\hat{\mathcal{J}}_1, \hat{W}_1'\}_{ij} \right), \\
V_{ij}^{\Gamma,3} &= -\frac{7\Gamma_0-2\Gamma_1}{360}\, \hat{\mathcal{J}}_{3i} \hattimes \hat{\mathcal{J}}_{3j}
  - \frac{3\Gamma_0+2\Gamma_1}{160}\, \hat{W}_{3i} \hattimes \hat{W}_{3j}
  - \frac{3\Gamma_0-2\Gamma_1}{96}\, \hat{W}_{3i}' \hattimes \hat{W}_{3j}'
  + \frac{\Gamma_0+\Gamma_1}{120}\, \{ \hat{\mathcal{J}}_3, \hat{W}_3 \}_{ij} \notag \\
&\quad{} - \frac{\Gamma_0}{16\sqrt{15}} \left(\{\hat{W}_1, \hat{W}_1'\}_{ij}
    - \{\hat{W}_2, \hat{W}_2'\}_{ij} \right)
  + \frac{\Gamma_0-5\Gamma_1}{48 \sqrt{15}} \left(\{\hat{\mathcal{J}}_1, \hat{W}_1'\}_{ij}
    - \{\hat{\mathcal{J}}_2, \hat{W}_2'\}_{ij} \right), \\
V_{ij}^{\Gamma,4} &= \frac{-5\Gamma_0+13\Gamma_1}{360}\, \{ \hat{\mathcal{J}}_2, \hat{\mathcal{J}}_3\}_{ij}
  + \frac{\Gamma_1}{80}\, \{\hat{W}_2, \hat{W}_3\}_{ij}
  - \frac{\Gamma_1}{48}\, \{ \hat{W}_2', \hat{W}_3' \}_{ij}
  + \frac{\Gamma_1}{16\sqrt{15}} \left(\{\hat{W}_2, \hat{W}_3'\}_{ij}
    - \{\hat{W}_3, \hat{W}_2'\}_{ij} \right) \notag \\
&\quad{} + \frac{5\Gamma_0-\Gamma_1}{240} \left(\{\hat{\mathcal{J}}_2, \hat{W}_3\}_{ij}
    + \{\hat{\mathcal{J}}_3, \hat{W}_2\}_{ij} \right)
  + \frac{5\Gamma_0-\Gamma_1}{48 \sqrt{15}} \left(\{\hat{\mathcal{J}}_2, \hat{W}_3'\}_{ij}
    - \{\hat{\mathcal{J}}_3, \hat{W}_2'\}_{ij} \right) \notag \\
&\quad{} + \frac{\Gamma_0+\Gamma_1}{8 \sqrt{15}}\, \{ \hat{W}_1, \hat{W}_7\}_{ij}
  - \frac{\Gamma_0+\Gamma_1}{24 \sqrt{15}}\, \{ \hat{\mathcal{J}}_1, \hat{W}_7\}_{ij}, \\
V_{ij}^{\Gamma,5} &= \frac{-5\Gamma_0+13\Gamma_1}{360}\, \{ \hat{\mathcal{J}}_3, \hat{\mathcal{J}}_1\}_{ij}
  + \frac{\Gamma_1}{80}\, \{\hat{W}_3, \hat{W}_1\}_{ij}
  - \frac{\Gamma_1}{48}\, \{ \hat{W}_3', \hat{W}_1' \}_{ij}
  + \frac{\Gamma_1}{16\sqrt{15}} \left(\{\hat{W}_3, \hat{W}_1'\}_{ij}
    - \{\hat{W}_1, \hat{W}_3'\}_{ij} \right) \notag \\
&\quad{} + \frac{5\Gamma_0-\Gamma_1}{240} \left(\{\hat{\mathcal{J}}_3, \hat{W}_1\}_{ij}
    + \{\hat{\mathcal{J}}_1, \hat{W}_3\}_{ij} \right)
  + \frac{5\Gamma_0-\Gamma_1}{48 \sqrt{3}} \left(\{\hat{\mathcal{J}}_3, \hat{W}_1'\}_{ij}
    - \{\hat{\mathcal{J}}_1, \hat{W}_3'\}_{ij} \right) \notag \\
&\quad{} + \frac{\Gamma_0+\Gamma_1}{8 \sqrt{15}}\, \{ \hat{W}_2, \hat{W}_7\}_{ij}
  - \frac{\Gamma_0+\Gamma_1}{24 \sqrt{15}}\, \{ \hat{\mathcal{J}}_2, \hat{W}_7\}_{ij}, \\
V_{ij}^{\Gamma,6} &= \frac{-5\Gamma_0+13\Gamma_1}{360}\, \{ \hat{\mathcal{J}}_1, \hat{\mathcal{J}}_2\}_{ij}
  + \frac{\Gamma_1}{80}\, \{\hat{W}_1, \hat{W}_2\}_{ij}
  - \frac{\Gamma_1}{48}\, \{ \hat{W}_1', \hat{W}_2' \}_{ij}
  + \frac{\Gamma_1}{16\sqrt{15}} \left(\{\hat{W}_1, \hat{W}_2'\}_{ij}
    - \{\hat{W}_2, \hat{W}_1'\}_{ij} \right) \notag \\
&\quad{} + \frac{5\Gamma_0-\Gamma_1}{240} \left(\{\hat{\mathcal{J}}_1, \hat{W}_2\}_{ij}
    + \{\hat{\mathcal{J}}_2, \hat{W}_1\}_{ij} \right)
  + \frac{5\Gamma_0-\Gamma_1}{48 \sqrt{3}} \left(\{\hat{\mathcal{J}}_1, \hat{W}_2'\}_{ij}
    - \{\hat{\mathcal{J}}_2, \hat{W}_1'\}_{ij} \right) \notag \\
&\quad{}+ \frac{\Gamma_0+\Gamma_1}{8 \sqrt{15}}\, \{ \hat{W}_3, \hat{W}_7\}_{ij}
  - \frac{\Gamma_0+\Gamma_1}{24 \sqrt{15}}\, \{ \hat{\mathcal{J}}_3, \hat{W}_7\}_{ij} .
\end{align}
\end{widetext}

\section{Effective interactions in the even-parity Cooper channel}
\label{sec:Cooperchannel_hlk}

We here demonstrate a decomposition of the interaction terms (\ref{eq:Vonsite}) and (\ref{eq:Vnn}) into the even-parity Cooper channels. The decomposition takes place through the generalized Fierz identity~\cite{Boettcher2017}
\begin{equation}
(\psi^{\dagger} N \psi)(\phi^{\dagger} M \phi) =\sum_{\hat{A},\hat{B}} f_{NM}(\hat{A},\hat{B})\,
  (\psi^{\dagger} \bar{A} \phi^{\dagger T})(\phi^{T} \bar{B}^{\dagger} \psi), \label{eq:Fierz_identity}
\end{equation}
with
\begin{equation}
f_{NM}(\hat{A},\hat{B}) = \frac{1}{16}\, \Tr ( U_T^{\dagger} \hat{A}N\hat{B}U_T M^T )
\label{eq:trace_formula}
\end{equation}
and
$\bar{A} \equiv \hat{A} U_T$, where $U_T$ is the unitary part of the time-reversal operator. In deriving Eq.~(\ref{eq:Fierz_identity}), we have used the orthogonality relation in Eq.\ (\ref{eq:traceorthog}).
This approach is useful for the construction of the effective interaction because the coefficients $f_{NM}(\hat A,\hat B)$ are given explicitly by the trace formula~(\ref{eq:trace_formula}).

In the following, we apply Eq.~(\ref{eq:Fierz_identity}) to the interaction terms and decompose them into the even-parity channels, i.e., $\hat{A}, \hat{B} \in \{ \mathbbm{1}, \hat{E}_1, \hat{E}_2, \hat{T}_1, \hat{T}_2, \hat{T}_3\}$.
The even-parity pairs satisfy $(\hat{A}U_T)^T = - \hat{A}U_T$ due to the Fermi statistics.
To this end, we first transform the interaction to momentum space and restrict it to pairing of electrons with opposite momenta,
\begin{align}
H_{\text{pair}} &= \frac{1}{2 N}\sum_{\mathbf{k},\mathbf{k}'}\sum_{\alpha,\beta,\alpha',\beta'}
  [V_{\mathbf{k},\mathbf{k}'}]_{\alpha \beta;\alpha' \beta'} \nonumber \\
&\quad{}\times c^\dagger_{\mathbf{k},\alpha}c^\dagger_{-\mathbf{k},\beta}
  c^{}_{{\bf -k'},\alpha'}c^{}_{{\bf k'},\beta'} .
\end{align}
The coupling strength contains contributions from the on-site interaction, Eq.\ (\ref{eq:Vonsite}), and from the nearest-neighbor interaction, Eq.\ (\ref{eq:Vnn}), as
\begin{equation}
V_{\mathbf{k},\mathbf{k}'} = V_{\mathbf{k},\mathbf{k}'}^o
  + V_{\mathbf{k},\mathbf{k}'}^e .
\end{equation}
From the trace formula (\ref{eq:trace_formula}), we obtain the on-site part
\begin{equation}
 V_{\mathbf{k},\mathbf{k}'}^o
  = \frac{U_0}{8}\, \bar{\mathbbm{1}} \bartimes \bar{\mathbbm{1}}
  + \frac{U_0}{24}\, \vec{\bar{T}} \bartimes \vec{\bar{T}},
\label{eq:CVonsite}
\end{equation}
where  the product $\bartimes$ is defined by, for a given field operator $c_{\mathbf{k}}^T \equiv (c_{\mathbf{k},\frac{3}{2}},c_{\mathbf{k},\frac{1}{2}},c_{\mathbf{k},-\frac{1}{2}},c_{\mathbf{k},-\frac{3}{2}})$,
\begin{align}
&\sum_{\alpha ,\beta, \alpha', \beta'}  (\bar{A} \bartimes \bar{B})_{\alpha \beta;\alpha' \beta' } \, c_{\mathbf{k},\alpha}^{\dagger} c_{-\mathbf{k}, \beta}^{\dagger} c_{-\mathbf{k}',\alpha'} c_{\mathbf{k}', \beta'} \nonumber \\
&\qquad \equiv \sum_{\alpha ,\beta, \alpha', \beta'} \bar{A}_{\alpha \beta} \bar{B}_{\beta'  \alpha' }^{\ast} \, c_{\mathbf{k},\alpha}^{\dagger} c_{-\mathbf{k}, \beta}^{\dagger} c_{-\mathbf{k}',\alpha'} c_{\mathbf{k}', \beta'} \nonumber \\
&\qquad = \left(\sum_{\alpha ,\beta}c_{\mathbf{k},\alpha}^{\dagger} \bar{A}_{\alpha \beta} c_{-\mathbf{k}, \beta}^{\dagger} \right) \left( \sum_{\alpha' ,\beta'} c_{-\mathbf{k}',\alpha'} \bar{B}_{\beta'  \alpha' }^{\ast} c_{\mathbf{k}', \beta'} \right) \nonumber \\
&\qquad =(c_{\mathbf{k}}^{\dagger} \bar{A} c_{-\mathbf{k}}^{\dagger T})(c_{-\mathbf{k}'}^T \bar{B}^{\dagger} c_{\mathbf{k}'}).
\end{align}
 If $\bar{A}$ and $\bar{B}$ are vectors of equal dimension, summation over their components is implied.

\begin{table*}
\caption{All even-parity nearest-neighbor pairing states and the corresponding irreps of the point group $O_h$. We adopt the abbreviations $c_\mu = \cos k_\mu a$, $s_\mu = \sin k_\mu a$. The symbols $\bar{\mathbbm{1}}$ etc.\ are defined in Table~\ref{tab:os}.  Entries that are nonzero at the $\Gamma$ point are marked in the rightmost column.}
\label{tab:nn}
\begin{ruledtabular}
\begin{tabular}{clc}
irrep & pairing state & nonzero at $\Gamma$ \\ \hline
$A_{1g}$ & $c_{A_{1g}} = (c_{y}c_z + c_xc_z + c_xc_y) \bar{\mathbbm{1}}$ & $\surd$ \\
& $c^{(E)}_{A_{1g}} = (c_{x}c_z - c_{y}c_z)\bar{E}_1
  + \frac{1}{\sqrt{3}}(c_yc_z + c_xc_z - 2c_xc_y)\bar{E}_2$ \\
& $s^{(T)}_{A_{1g}} = s_ys_z\bar{T}_1 + s_xs_z\bar{T}_2 + s_xs_y\bar{T}_3$ \\
$A_{2g}$ & $c^{(E)}_{A_{2g}} = (c_xc_z - c_yc_z)\bar{E}_2
  - \frac{1}{\sqrt{3}}(c_yc_z + c_xc_z - 2c_xc_y)\bar{E}_1$ \\
$E_g$ & $\vec{c}_{E_g} = \left( c_xc_z - c_yc_z,
  \frac{1}{\sqrt{3}}(c_{y}c_z + c_xc_z - 2c_xc_y)\right) \bar{\mathbbm{1}}$ \\
& $\vec{c}^{\,(E)}_{E_g} = (c_{y}c_z + c_xc_z + c_xc_y)\left( \bar{E}_1,\bar{E}_2 \right)$ & $\surd$ \\
& $\vec{\tilde{c}}^{\,(E)}_{E_g}
  = \left( (c_xc_z - c_yc_z)\bar{E}_2 + \frac{1}{\sqrt{3}}(c_yc_z + c_xc_z - 2c_xc_y)\bar{E}_1,
  (c_xc_z - c_yc_z)\bar{E}_1 - \frac{1}{\sqrt{3}}(c_yc_z + c_xc_z - 2c_xc_y)\bar{E}_2\right)$ \\
& $\vec{s}^{\,(T)}_{E_g} = \left(s_ys_z\bar{T}_1 - s_xs_z\bar{T}_2,
  \frac{1}{\sqrt{3}}(2s_xs_y\bar{T}_3 - s_ys_z\bar{T}_1 - s_xs_z\bar{T}_2)\right)$ \\
$T_{1g}$ & $\vec{c}^{\,(T)}_{T_{1g}}
  = \left ((c_{x}c_z - c_xc_y)\bar{T}_1,\,(c_{x}c_y - c_yc_z)\bar{T}_2,
  (c_{y}c_z - c_xc_z)\bar{T}_3 \right)$ \\
& $\vec{s}^{\,(E)}_{T_{1g}}
  = \left( \frac{1}{2}s_{y}s_z(\sqrt{3}\bar{E}_2 + \bar{E}_1),
  \frac{1}{2}s_{x}s_z(\sqrt{3}\bar{E}_2 - \bar{E}_1),
  s_{x}s_y\bar{E}_1 \right)$ \\
& $\vec{s}^{\,(T)}_{T_{1g}}
  = \left((s_{x}s_y\bar{T}_2-s_xs_z\bar{T}_3),
  (s_ys_z\bar{T}_3-s_xs_y\bar{T}_1),
  (s_xs_z\bar{T}_1-s_ys_z\bar{T}_2) \right)$ \\
$T_{2g}$ & $\vec{c}^{\,(T)}_{T_{2g}}
  = (c_{y}c_z + c_xc_z + c_xc_y)\left( \bar{T}_1, \bar{T}_2, \bar{T}_3 \right)$ & $\surd$ \\
& $\vec{\tilde{c}}^{\,(T)}_{T_{2g}}
  = \left( (c_xc_y + c_xc_z - 2c_yc_z)\bar{T}_1,
  (c_xc_y + c_yc_z - 2c_xc_z)\bar{T}_2,
  (c_yc_z + c_xc_z - 2c_xc_y)\bar{T}_3 \right)$ \\
& $\vec{s}_{T_{2g}} = (s_ys_z, s_xs_z, s_xs_y) \bar{\mathbbm{1}}$ \\
& $\vec{s}^{\,(E)}_{T_{2g}}
  = \left( \frac{1}{2}s_{y}s_z(\sqrt{3}\bar{E}_1 -\bar{E}_2),
  - \frac{1}{2}s_{x}s_z(\sqrt{3}\bar{E}_1 +\bar{E}_2),
  s_{x}s_y\bar{E}_2 \right)$ \\
& $\vec{s}^{\,(T)}_{T_{2g}} = \left( (s_{x}s_y\bar{T}_2+s_xs_z\bar{T}_3),
  (s_ys_z\bar{T}_3+s_xs_y\bar{T}_1), (s_xs_z\bar{T}_1+s_ys_z\bar{T}_2) \right)$ \\
\end{tabular}
\end{ruledtabular}
\end{table*}

The coefficients of the nearest-neighbor interaction are also determined from the trace formula; the result comprises extended \textit{s}-wave and \textit{d}-wave channels. For instance, the charge-charge interaction is decomposed into Cooper channels in terms of irreps of $O_h$ as
\begin{widetext}
\begin{align}
 V_{\mathbf{k},\mathbf{k}'}^{e,\, U_1} &= \frac{U_1}{18}\, c_{A_{1g}} \bartimes c_{A_{1g}}'
  + \frac{U_1}{12}\, c_{A_{1g}}^{(E)} \bartimes c_{A_{1g}}^{(E) \prime}
  + \frac{U_1}{6}\, s_{A_{1g}}^{(T)} \bartimes s_{A_{1g}}^{(T) \prime}
  - \frac{U_1}{6 \sqrt{3}} \left( c_{A_{1g}} \bartimes s_{A_{1g}}^{(T) \prime}
    +  s_{A_{1g}}^{(T) } \bartimes c_{A_{1g}}' \right)
  + \frac{U_1}{12}\, c_{A_{2g}}^{\;(E)} \bartimes c_{A_{2g}}^{\;(E)\prime} \notag \\
&\quad{}+ \frac{U_1}{12}\, \vec{c}_{E_g} \bartimes \vec{c}_{E_g}^{\; \prime}
  + \frac{U_1}{9}\, \vec{c}_{E_g}^{\; (E)} \bartimes \vec{c}_{E_g}^{\; (E) \prime}
  + \frac{U_1}{12}\, \vec{\tilde{c}}_{E_g}^{\; (E)} \bartimes \vec{\tilde{c}}_{E_g}^{\; (E) \prime}
  + \frac{U_1}{4}\, \vec{s}_{E_g}^{\; (T)} \bartimes \vec{s}_{E_g}^{\; (T) \prime}
  + \frac{U_1}{4 \sqrt{3}} \left( \vec{c}_{E_g} \bartimes \vec{s}_{E_g}^{\; (T) \prime}
    + \vec{s}_{E_g}^{\; (T)} \bartimes \vec{c}_{E_g}^{\; \prime} \right) \notag \\
&\quad{}+ \frac{U_1}{12} \left(\vec{c}_{T_{1g}}^{\;(T) }
    + \vec{s}_{T_{1g}}^{\;(T) }\right) \bartimes \left(\vec{c}_{T_{1g}}^{\;(T) \prime}
    + \vec{s}_{T_{1g}}^{\;(T) \prime }\right)
  + \frac{U_1}{3}\, \vec{s}_{T_{1g}}^{\; (E)} \bartimes \vec{s}_{T_{1g}}^{\; (E) \prime}
  + \frac{5U_1}{54}\, \vec{c}_{T_{2g}}^{\; (T)} \bartimes \vec{c}_{T_{2g}}^{\; (T) \prime}
  + \frac{7 U_1}{108}\, \vec{\tilde{c}}_{T_{2g}}^{\; (T)} \bartimes \vec{\tilde{c}}_{T_{2g}}^{\; (T)} \notag \\
&\quad{}+ \frac{U_1}{6}\, \vec{s}_{T_{2g}} \bartimes \vec{s}_{T_{2g}}^{\, \prime}
  + \frac{U_1}{3}\, \vec{s}_{T_{2g}}^{\; (E)} \bartimes \vec{s}_{T_{2g}}^{\; (E) \prime}
  + \frac{U_1}{12}\, \vec{s}_{T_{2g}}^{\; (T)} \bartimes \vec{s}_{T_{2g}}^{\; (T) \prime}
  - \frac{U_1}{27} \left(\vec{c}_{T_{2g}}^{\;(T)} \bartimes \vec{\tilde{c}}_{T_{2g}}^{\; (T) \prime}
    + \vec{\tilde{c}}_{T_{2g}}^{\;(T)} \bartimes \vec{c}_{T_{2g}}^{\; (T) \prime} \right) \notag \\
&\quad{}- \frac{U_1}{6 \sqrt{3}} \left( \vec{c}_{T_{2g}}^{\;(T)} \bartimes \vec{s}_{T_{2g}}^{\; \prime}
    + \vec{s}_{T_{2g}} \bartimes \vec{c}_{T_{2g}}^{\; (T) \prime}\right)
  - \frac{U_1}{18} \left( \vec{c}_{T_{2g}}^{\;(T)} \bartimes \vec{s}_{T_{2g}}^{\; (T) \prime}
    + \vec{s}_{T_{2g}}^{\; (T)} \bartimes \vec{c}_{T_{2g}}^{\; (T) \prime}\right)
  + \frac{U_1}{6 \sqrt{3}} \left( \vec{\tilde{c}}_{T_{2g}}^{\;(T)} \bartimes \vec{s}_{T_{2g}}^{\;\prime}
    + \vec{s}_{T_{2g}} \bartimes \vec{\tilde{c}}_{T_{2g}}^{\; (T) \prime}\right) \notag \\
&\quad{}- \frac{U_1}{36} \left( \vec{\tilde{c}}_{T_{2g}}^{\;(T)} \bartimes \vec{s}_{T_{2g}}^{\; (T)\prime}
    + \vec{s}_{T_{2g}}^{\; (T)} \bartimes \vec{\tilde{c}}_{T_{2g}}^{\; (T) \prime}\right), \label{eq:NNint}
\end{align}
\end{widetext}
where the representations of pairing states (matrix-valued functions) are tabulated in Table~\ref{tab:nn} and the prime refers to the primed momentum coordinates. A general analysis taking into account all contributions to the interaction and all pairing channels would be extremely laborious. However, we should bear in mind that the projection to the $j = 3/2$ subspace is only valid close to the $\Gamma$ point. Most of the states tabulated in Table~\ref{tab:nn} are quadratic in $\mathbf{k}$ close to $\Gamma$ (i.e., \textit{d}-wave like); the exceptions are the three states marked in Table\ \ref{tab:nn}, which correspond to the extended \textit{s}-wave form factor, and which have a finite value at the $\Gamma$ point. Since the extended \textit{s}-wave states have similar coupling constants compared to the \textit{d}-wave states, we expect that for sufficiently small chemical potential relative to the band touching point the extended \textit{s}-wave states will be the leading instabilities since the \textit{d}-wave states will open a much smaller gap at the Fermi surface. We similarly expect that \textit{p}-wave states in the odd parity channel will not be leading instabilities. We can thus ignore the \textit{d}-wave states and focus upon the \textit{s}-wave states, and so approximate the pairing interaction from the charge-charge coupling as
\begin{align}
 V_{\mathbf{k},\mathbf{k}'}^{e,\, U_1} &\approx \frac{U_1}{18}\, c_{A_{1g}} \bartimes c_{A_{1g}}'
   + \frac{U_1}{9}\, \vec{c}_{E_g}^{\; (E)} \bartimes \vec{c}_{E_g}^{\; (E) \prime} \notag \\
&\quad{} + \frac{5U_1}{54}\, \vec{c}_{T_{2g}}^{\; (T)} \bartimes \vec{c}_{T_{2g}}^{\; (T) \prime}.
\label{eq:extend-s_cc}
\end{align}
For the same reason, we ignore the \textit{d}-wave states for the spin interactions. Using the same procedure, the pairing interaction from the spin coupling is obtained as
\begin{align}
V_{\mathbf{k},\mathbf{k}'}^{e,\,\text{spin}} &\approx \left(-\frac{J}{216} - \frac{D}{27}
  - \frac{\Gamma_0-2\Gamma_1}{108} \right) c_{A_{1g}} \bartimes c_{A_{1g}}' \notag \\
&\quad{}+ \left(- \frac{J}{108} +\frac{D}{27}  - \frac{\Gamma_0+\Gamma_1}{54} \right)
  \vec{c}_{E_g}^{\; (E)} \bartimes \vec{c}_{E_g}^{\; (E) \prime} \notag \\
&\quad{}+ \left(\frac{J}{216} - \frac{D}{27} +\frac{\Gamma_0+2\Gamma_1}{108} \right)
  \vec{c}_{T_{2g}}^{\; (T)} \bartimes \vec{c}_{T_{2g}}^{\; (T) \prime}. \label{eq:extend-s_spin}
\end{align} 
Equations (\ref{eq:CVonsite}), (\ref{eq:extend-s_cc}), and (\ref{eq:extend-s_spin}) correspond to Eq.~(\ref{eq:pair_int}).

\section{Details of numerical solution of the gap equation}
\label{sec:BCS}

In this Appendix, we provide some background on the numerical solution of the BCS gap equation for the $E_g$ order parameter $(\Delta_1,\Delta_2)$. A more detailed discussion will be given in a future work \cite{BhT22}. Both the gap equation (\ref{eq:BCSgapeq}) and Eq.\ (\ref{eq:BCSfreeen}) for the internal energy involve integration over the three-dimensional Brillouin zone, which is the main complication compared to the textbook calculation for a parabolic band. In Eq.\ (\ref{eq:BCSfreeen}), we take the difference of the momentum contributions to the internal energy in the normal state and in the superconducting state first and then perform the integral to get $F_n-F_s$ plotted in Fig.\ \ref{fig.BCS1}. This strongly reduces round-off errors.

The main problem for accurate numerics then stems from the form of the integrand. We here discuss the case of Eq.\ (\ref{eq:BCSfreeen}), the situation for Eq.\ (\ref{eq:BCSgapeq}) is analogous. For a simple parabolic band and constant pairing amplitude $\Delta$, the integrand is proportional to
\begin{equation}
\delta\epsilon_\mathbf{k} = \sqrt{\xi_\mathbf{k}^2 + \Delta^2} - \xi_\mathbf{k} ,
\label{eq.BCS.De}
\end{equation}
where $\xi_\mathbf{k}$ is the normal-state dispersion relative to the chemical potential. (In our case the expression is more complicated but the essential points remain.) The radial integral diverges logarithmically at large momenta $k$. The integral is cut off at large $k$ corresponding to an energy scale $\Lambda$, leading to a term proportional to $\ln(\Delta/\Lambda)$. The appearance of the large scale $\Lambda$ and the small scale $\Delta$ shows that the integral is sensitive to the whole of momentum space. For our lattice model, the integral is naturally cut off by the finite Brillouin zone but still the full Brillouin zone is important for accurate results.

We perform the integrals using spherical coordinates. The radial integral is performed first, inside the angular integrals. From Eq.\ (\ref{eq.BCS.De}), we expect that momenta close to the normal-state Fermi momentum $k_F$ will contribute most and, since $\xi_\mathbf{k}$ is linear in $k$, the integrand changes on a momentum scale proportional to $\Delta$. Therefore, we split the radial integral into four parts $[0,k_F-k_1]$, $[k_F-k_1,k_F]$, $[k_F,k_F+k_2]$, and $[k_F+k_2,k_\mathrm{BZ}(\theta,\phi)]$, where $k_1$ and $k_2$ are proportional to $\Delta$ at $k_F$ and $k_\mathrm{BZ}(\theta,\phi)$ describes the surface of the Brillouin zone in the direction $\theta$, $\phi$. The constants of proportionality in $k_1$ and $k_2$ are chosen so as to minimize numerical noise. The integrals are performed using globally adaptive sampling as implemented in Mathematica (version 12) with the accuracy goal typically set to 18 digits and the maximum number of recursions set to 12 for the two outer intervals and to 8 for the two inner intervals.

The resulting integrand for the wrapping integrals over angles $\theta$ and $\phi$ is a well-behaved function. For these integrals, we also use globally adaptive sampling, with the accuracy goal set to 18 digits and the maximum number of recursions set to 4.

The main diagnostics for the quality of the numerical integration are (a) the observation that it gives smooth $F_n-F_s$ and also $\Delta$ (not shown) vs.\ $V_0$ down to very small $F_n-F_s$ and $\Delta$ and (b) that the results in this range agree with the expected scaling for weak-coupling BCS theory. The numerical noise is small compared to the thickness of the lines in Fig.\ 2. Also note that the crossings of lines in Figs.\ 2(a) and (b) take place in a range where $\Delta$ and $F_n-F_s$ are so large that the numerical integration is unproblematic in any case. The BCS scaling results from the leading terms in the energy difference being
\begin{equation}
F_s-F_n = a\, \Delta^2 \ln \frac{\Delta}{\Lambda} + b\, \Delta^2 + c\, \frac{\Delta^2}{V_0} ,
\end{equation}
where $a$, $b$, $c$ are constants. The first two terms are due to the quasiparticle contribution, whereas the third stems from the mean-field decoupling. Minimization with respect to $\Delta$ gives the BCS results $\Delta \sim e^{-c/aV_0}$ and $F_s-F_n \sim - e^{-2c/aV_0}$. This leads to
\begin{equation}
\ln \frac{F_n-F_s}{t_\pi} \cong \mathrm{const} - \frac{2c}{aV_0} ,
\end{equation}
which is seen in Fig.~2(a).

\end{document}